\documentclass[preprintnumbers,prd,twocolumn,amsmath,amssymb,nofootinbib,superscriptaddress]{revtex4-2}

\usepackage{graphicx}
\usepackage{color}
\usepackage{bm}

\newcommand{\beq}{\begin{equation}}
\newcommand{\eeq}{\end{equation}}
\newcommand{\bea}{\begin{eqnarray}}
\newcommand{\eea}{\end{eqnarray}}
\newcommand{\bseq}{\begin{subequations}}
\newcommand{\eseq}{\end{subequations}}

\newcommand{\rf}[1]{(\ref{#1})}

\begin{document}

\title{Cosmological models with arbitrary spatial curvature in the theory of gravity with non-minimal derivative coupling}

\author{Sergey V. Sushkov} 
\email{sergey_sushkov@mail.ru}
\affiliation{Institute of Physics, Kazan Federal University, Kremliovskaya street 18, Kazan 420008, Russia}

\author{Rafkat Galeev}
\email{rafgaleev3@gmail.com}
\affiliation{Institute of Physics, Kazan Federal University, Kremliovskaya street 18, Kazan 420008, Russia}

\begin{abstract}
We investigate isotropic and homogeneous cosmological scenarios in the scalar-tensor theory of gravity with non-minimal derivative coupling of a scalar field to the curvature given by the term $(\zeta/H_0^2) G^{\mu\nu}\nabla_\mu\phi \nabla_\nu\phi$ in the Lagrangian. 
In general, a cosmological model is determined by six dimensionless parameters: the coupling parameter $\zeta$, and density parameters $\Omega_0$ (cosmological constant), $\Omega_2$ (spatial curvature term), $\Omega_3$ (non-relativistic matter), $\Omega_4$ (radiation), $\Omega_6$ (scalar field term) (see Eqs. \rf{param1}, \rf{param2}), and the universe evolution is described by the modified Friedmann equation \rf{geneqH}. In the case $\zeta=0$ (no non-minimal derivative coupling) and $\Omega_6=0$ (no scalar field) one has the standard $\Lambda$CDM-model, while if $\Omega_6\not=0$ -- the $\Lambda$CDM-model with an ordinary scalar field. As is well-known, this model has an initial singularity, the same for all $k$ ($k=0,\pm1$), while its global behavior depends on $k$. The universe expands eternally if $k=0$ (zero spatial curvature) or $k=-1$ (negative spatial curvature), while in case $k=+1$ (positive spatial curvature) the universe expansion is changed to contraction, which is ended by a final singularity. 
The situation is {\em crucially} changed when the scalar field possesses non-minimal derivative coupling to the curvature, i.e. when $\zeta\not=0$. 
Now, depending on model parameters, (i) There are three qualitatively different initial state of the universe: an {\em eternal kinetic inflation}, an {\em initial singularity}, and a {\em bounce}. The bounce is possible for {\em all} types of spatial geometry of the homogeneous universe; (ii)  For {\em all} types of spatial geometry, the universe goes inevitably through the {\em primary quasi-de Sitter} (inflationary) epoch when $a(t)\propto e^{h_{dS}(H_0t)} $ with the de Sitter parameter $h_{dS}^2=
{1}/{9\zeta}-{8\zeta\Omega_2^3}/{27\Omega_6}$.
The mechanism of primary or {\em kinetic} inflation is provided by non-minimal derivative coupling and needs no fine-tuned potential; (iii) There are {\em cyclic} scenarios of the universe evolution with the non-singular bounce at a minimal value of the scale factor, and a turning point at the maximal one; (iv) There is a natural mechanism providing a {\em change} of cosmological epochs.  
%

\end{abstract}

 \maketitle

\section{Introduction}
In recent decades the observational cosmology has been going through the period of the rapid growth. Precise measurements of the Cosmic Microwave Background (CMB) radiation \cite{CMB}, systematic observations of nearby and distant Type Ia supernovae (SNe Ia) \cite{supernova}, study of  baryon acoustic oscillations \cite{BAO}, mapping the large-scale structure of the Universe, microlensing observations, and many other remarkable results (see, for example, the review \cite{Observations}) have essentially expanded our knowledge about the Universe. Amazing discoveries, such as the accelerating expansion of the Universe and the dark matter evidence, have set new serious challenges before theoretical cosmology faced the necessity of radical modification of the standard model having successfully been exploited for a long time. Now, any viable cosmological model has to be able to describe several qualitatively different epoches of the Universe evolution, including the primary inflation, the matter-dominated stage, and the present acceleration (or secondary inflation). Moreover, it should also explain a mechanism providing an epoch change. These challenges have prompted many speculations mostly based on phenomenological ideas which involve new dynamical sources of gravity that act as dark energy, and/or various modifications to general relativity. To date, many different versions of modified or extended theories of gravity have been proposed (see surveys 
\cite{Review_Salvatore:2011, Review_Clifton_etal, Review_ModGrav:2013, Review_Berti_etal, Review_Nojiri:2017, Review_Langlois:2019} and references therein).\footnote{This plethora of models reflects a deep crisis of phenomenological approach in the modern theoretical cosmology. To date, there are no unique criteria  to prefer one or another phenomenological model.} One of such models intensively studied today is Horndeski theory of gravity \cite{Horndeski} derived in the 1970s as an attempt to obtain the most general action for a scalar-tensor theory with a single scalar degree of freedom and second-order field equations. In 2011 Horndeski gravity has been rediscovered in the context of generalized Galileon theories \cite{Kobayashi:2011}, and since the interest in this model has only growing.\footnote{The literature dedicated to various aspects of Horndeski gravity is very vast, and its survey lays out of the scope of this work. The reader interesting in this topic can find some references in the already mentioned surveys \cite{Review_Berti_etal, Review_Clifton_etal}.}

The important subclass of Horndeski gravity is represented by models with a non-minimal derivative coupling of a scalar field $\phi$ with the Einstein tensor with the action
\bea
	S &=& \displaystyle\frac12\int d^4x\sqrt{-g}\,
	\bigg[\frac{1}{8\pi} (R-\Lambda) 
	\nonumber\\
	&& -\left( g^{\mu\nu}+\eta G^{\mu\nu} \right)\nabla_{\mu}\phi\nabla_{\nu}\phi\bigg]+S^{(m)},
	\label{action}
\eea
where $R$ is the scalar curvature, $G_{\mu\nu}$ is the Einstein tensor, $\Lambda$ is the cosmological constant, $\eta$ is the coupling parameter, 
and $S^{(m)}$ is the action for ordinary matter fields, supposed to be minimally coupled to gravity in the usual way.

The theory of gravity with non-minimal derivative coupling involves the additional dimensional parameter $\eta$ with dimension of ({\em length})$^2$, which leads to  interesting features of astrophysical objects. In particular, black holes \cite{Rinaldi:2012, Minamitsuji:2013, Anabalon:2014, Babichev:2014, Kobayashi:2014, Babichev:2015}, wormholes \cite{Sushkov:2012b, Sushkov:2014}, and neutron stars \cite{Rinaldi:2015, Rinaldi:2016, Silva:2016, Maselli:2016, Eickhoff:2018, KasSus:2023} have been widely explored within this theory.   		
As well, the non-minimal derivative coupling leads to very interesting cosmological consequences, which have been intensively studied in our recent works
\cite{Sushkov:2009, SarSus:2010, Sushkov:2012, SkuSusTop:2013, MatSus:2015, StaSusVol:2016, StaSusVol:2019}.
The most important feature we have found is that the non-minimal derivative coupling provides an essentially new inflationary mechanism and naturally describes transitions between various cosmological phases without any fine-tuned potential \cite{Sushkov:2009, SarSus:2010, Sushkov:2012}. The inflation is driving by terms in the field equations responsible for the non-minimal derivative coupling. At early times these terms are dominating, and the cosmological evolution has the quasi-de Sitter character $a(t)\sim e^{H_\eta t}$ with $H_\eta=1/\sqrt{9\eta}$. Later on, in the course of cosmological evolution the domination of $\eta$-terms is canceled, and this leads to a change of cosmological epochs. More generally, the above-mentioned features have been reopened in Ref. \cite{StaSusVol:2016} as a part of screening mechanism providing the screening of the $\Lambda$-term and matter at early times of universe evolution. Surprisingly, in Ref. \cite{StaSusVol:2019} we find that the same mechanism provides the screening of anisotropies at early time within the Bianchi I homogeneous spacetime model. Therefore, contrary to what one would normally expect, the early state of the universe in the theory of gravity with non-minimal derivative coupling cannot be anisotropic in the absence of spatial curvature.

It is worth noting that the most of results given in \cite{Sushkov:2009, SarSus:2010, Sushkov:2012, SkuSusTop:2013, MatSus:2015, StaSusVol:2016, StaSusVol:2019} and mentioned above have been obtained for cosmological models with zero spatial curvature. At the same time, it is well-know that the nonzero spatial curvature can essentially change a character of the universe evolution. Some preliminary results, taking into account the spatial curvature, have been obtained in \cite{StaSusVol:2016}, where a systematic analysis of homogeneous and isotropic cosmologies in the theory of gravity with non-minimal derivative coupling had been represented. In Ref. \cite{StaSusVol:2016} we analyzed a rich spectrum of solutions focusing mostly on their asymptotic properties, while the global solutions describing the entire evolution of the universe had been analyzed only briefly for the case of zero spatial curvature. 

In this work we explore in details both asymptotic and global homogeneous and isotropic cosmological solutions in the theory \rf{action} in models containing also a $\Lambda$-term and a matter.     

The paper is organized as follows. Equations describing homogeneous
and isotropic cosmologies in the theory \rf{action} are derived in the next section \ref{II}. 
Solutions of these equations are constructed and analyzed in Sec. III first in the early and late time limits and then globally. In subsections of Sec. III we consider several models with different sets of cosmological parameters, starting with the simplest model which contains only an ordinary scalar field, and finishing the most general model with the scalar field possessing the non-minimal derivative coupling with the curvature, cosmological constant $\Lambda$, and matter. In the last section we summarize obtained results.

\section{Field equations\label{II}}
Varying the action \rf{action} with respect to $g_{\mu\nu}$ and $\phi$ gives the field equations, respectively:
\bseq\label{fieldeq}
\bea
\label{eineq}
&& G_{\mu\nu}=-g_{\mu\nu}\Lambda+8\pi\big[T_{\mu\nu}^{(m)}+T_{\mu\nu}^{(\phi)}
+\eta \Theta_{\mu\nu}\big], \\
\label{eqmo}
&&[g^{\mu\nu}+\eta G^{\mu\nu}]\nabla_{\mu}\nabla_\nu\phi=0,
\eea
\eseq
where $T^{(m)}_{\mu\nu}$ is a stress-energy
tensor of ordinary matter, and
\bea \label{T}
T^{(\phi)}_{\mu\nu}&=&\nabla_\mu\phi\nabla_\nu\phi-
{\textstyle\frac12}g_{\mu\nu}(\nabla\phi)^2, \\
\Theta_{\mu\nu}&=&-{\textstyle\frac12}\nabla_\mu\phi\,\nabla_\nu\phi\,R
+2\nabla_\alpha\phi\,\nabla_{(\mu}\phi R^\alpha_{\nu)}
\nonumber\\
&&
+\nabla^\alpha\phi\,\nabla^\beta\phi\,R_{\mu\alpha\nu\beta}
+\nabla_\mu\nabla^\alpha\phi\,\nabla_\nu\nabla_\alpha\phi
\nonumber\\
&&
-\nabla_\mu\nabla_\nu\phi\,\square\phi-{\textstyle\frac12}(\nabla\phi)^2
G_{\mu\nu}
\label{Theta}\\
&&
+g_{\mu\nu}\big[-{\textstyle\frac12}\nabla^\alpha\nabla^\beta\phi\,
\nabla_\alpha\nabla_\beta\phi
+{\textstyle\frac12}(\square\phi)^2
\nonumber\\
&& \ \ \ \ \ \ \ \ \ \ \ \ \ \ \ \ \ \  \ \ \   \ \ \  \ \ \ \ \ \
\ \ \ \ \ -\nabla_\alpha\phi\,\nabla_\beta\phi\,R^{\alpha\beta}
\big]. \nonumber
\eea
Due to Bianchi identity $\nabla^\mu G_{\mu\nu}=0$ and the conservation law
$\nabla^\mu T^{(m)}_{\mu\nu}=0$, Eq. \rf{eineq} leads to the differential
consequence
\beq
\label{BianchiT}
\nabla^\mu\big[T^{(\phi)}_{\mu\nu}+\eta \Theta_{\mu\nu}\big]=0.
\eeq
Substituting Eqs. \rf{T} and \rf{Theta} into \rf{BianchiT}, one can check straightforwardly that the differential consequence \rf{BianchiT} is equivalent to \rf{eqmo}.  In other words, Eq. \rf{eqmo} is a differential consequence of Eq. \rf{eineq}.

Let us consider FRW cosmological models with the metric
\begin{equation}
\label{metric} 
ds^2=-dt^2+\textrm{a}^2(t)\left[\frac{dr^2}{1-kr^2}+r^2(d\theta^2 +\sin^2\theta d\varphi^2)\right],
\end{equation}
where $k=0,\pm1$, $\textrm{a}(t)$ is the scale factor, and $H(t)=\dot{\textrm{a}}(t)/\textrm{a}(t)$ is the Hubble parameter. 
Denoting the present moment of time as $t_0$, we have $\textrm{a}_0=\textrm{a}(t_0)$ and $H_0=H(t_0)$. 
Supposing homogeneity and isotropy, we also get $\phi=\phi(t)$ and $T^{(m)}_{\mu\nu}=\mathop{\rm diag}(\rho,p,p,p)$, where $\rho=\rho(t)$ is the energy density and $p=p(t)$ is the pressure of matter.


The general field equations \rf{fieldeq} written for the metric \rf{metric}
give the following two independent equations:
\bseq\label{genfieldeq}
\bea
  \label{00cmpt}
  &&3\left(H^2+\frac{k}{\textrm{a}^2}\right) =\Lambda +8\pi\rho
  \nonumber\\
  &&\qquad\qquad\qquad\quad
    +4\pi{\psi}^2\left(1-9\eta \left(H^2+\frac{k}{3\textrm{a}^2}\right)\right),
  \\
  \label{eqmocosm}
  &&\psi\left(1-3\eta \left(H^2+\frac{k}{\textrm{a}^2}\right)\right)=\frac{Q}{\textrm{a}^3}, 
\eea
\eseq
where we denote $\psi=\dot\phi$. 
Here Eq. \rf{00cmpt} is the modified Friedmann equation, i.e. the $tt$-component of \rf{eineq}, while Eq. \rf{eqmocosm} is the first integral of the scalar field equation \rf{eqmo}, where $Q$ is a constant of integration.

Assume that the matter filling the universe is a mixture of a radiation and a non-relativistic component:
\beq\label{matter}
\rho=\rho_m+\rho_r=\rho_{{m}0} \left(\frac{\textrm{a}_0}{\textrm{a}}\right)^{3}
+\rho_{{r}0} \left(\frac{\textrm{a}_0}{\textrm{a}}\right)^{4}.
\eeq


Now let us introduce the dimensionless scale factor $a$, Hubble parameter $h$, and coupling parameter $\zeta$ as follows: 
\beq\label{param1}
a=\frac{\textrm{a}}{\textrm{a}_0}, \quad 
h=\frac{H}{H_0}, \quad
\zeta=\eta H_0^2,
\eeq
and the following dimensionless density parameters:
$$
\Omega_{0}=\frac{\Lambda}{3H_0^2}, \quad
\Omega_{2}=\frac{k}{\textrm{a}_0^2 H_0^2}, \quad
\Omega_{3}=\frac{\rho_{m0}}{\rho_{cr}}, \quad
$$
\beq\label{param2}
\Omega_{4}=\frac{\rho_{r0}}{\rho_{cr}}, \quad
\Omega_{6}=\frac{4\pi Q^2}{3\textrm{a}_0^6 H_0^2},
\eeq
where $\rho_{cr}=3H_0^2/8\pi$ is the critical density. We assume in this work that $\Lambda \ge 0$, hence $\Omega_0$ is always not negative, i.e. $\Omega_0\ge 0$, and the sign of $\Omega_2$ is
the same as that of $k$. Here it is also worth to emphasize the physical meaning of the dimensionless coupling parameter $\zeta$. 
The parameter $\eta$ has the dimension $(length)^2$, and so it will be convenient to use the notation $\eta=\varepsilon \ell^2$, where $\varepsilon$ is the sign of $\eta$, i.e. $\varepsilon=\pm 1$, and $\ell$ is a characteristic length which characterizes the nonminimal derivative coupling between the scalar field and curvature. The value $H_0$ determines the size of Hubble horizon as 
$ 
\ell_H={1}/{H_0}.
$ 
Hence, $\zeta$ is proportional to the square of ratio of two characteristic scales:
\beq
\zeta=\varepsilon \left(\frac{\ell}{\ell_H}\right)^2.
\eeq

Now, substituting $\psi$ from Eq. \rf{eqmocosm} into \rf{00cmpt}, we can rewrite the modified Friedmann equation in terms of dimensionless values:
\beq\label{geneqH}
h^2=\Omega_{0}
-\frac{\Omega_2}{a^2} 
+\frac{\Omega_3}{a^3} 
+\frac{\Omega_4}{a^4}
+\frac{\Omega_{6}\big(1-3\zeta (3h^2+\frac{\Omega_2}{a^2})\big)}
{a^6 \big(1-3\zeta (h^2 +\frac{\Omega_2}{a^2})\big)^2}.
\eeq
Denoting $y=h^2$ and bringing all terms in \rf{geneqH} to the common denominator
yields
\beq\label{cube}
\frac{P(a,y)}{\big(1-3\zeta (y +\frac{\Omega_2}{a^2})\big)^2}=0,
\eeq
where
\beq
P(a,y)=y^3+c_2(a)y^2+c_1(a)y+c_0(a)
\eeq
is the cubic in $y$ polynomial with the coefficients
\bea
c_2 &=& -\Omega_0+\frac{3\Omega_2}{a^2}-\frac{\Omega_3}{a^3}-\frac{\Omega_4}{a^4}
-\frac{2}{3\zeta},
\nonumber\\
c_1 &=& -\frac{2\Omega_2}{a^2}\left(\Omega_0-\frac32 \frac{\Omega_2}{a^2} +\frac{\Omega_3}{a^3}+\frac{\Omega_4}{a^4}\right)
\nonumber\\
&&
+\frac{1}{3\zeta}\left(2 \Omega_0-\frac{4\Omega_2}{a^2} +\frac{2\Omega_3}{a^3}+\frac{2\Omega_4}{a^4} +\frac{3\Omega_6}{a^6}\right)
+\frac{1}{9\zeta^2},
\nonumber\\
c_0 &=& -\frac{\Omega_2^2}{a^4} \left(\Omega_0-\frac{\Omega_2}{a^2} +\frac{\Omega_3}{a^3}+\frac{\Omega_4}{a^4}\right)
\nonumber\\
&&
+\frac{\Omega_2}{3a^2\zeta} \left(2\Omega_0 -\frac{2\Omega_2}{a^2} +\frac{2\Omega_3}{a^3} +\frac{2\Omega_4}{a^4} +\frac{\Omega_6}{a^6}\right)
\nonumber\\
&&
-\frac{1}{9\zeta^2} \left(\Omega_0-\frac{\Omega_2}{a^2} +\frac{\Omega_3}{a^3} +\frac{\Omega_4}{a^4} +\frac{\Omega_6}{a^6}\right).
\eea
Eq.\rf{cube} will be fulfilled if $P(a, y) = 0$ and $1-3\zeta (y +\frac{\Omega_2}{a^2})\not= 0$, hence the problem reduces to studying roots of the cubic polynomial. Finding a particular root $y_i$, one determines the algebraic dependence of the Hubble parameter $h$ on the scale factor $a$. The relation to the physical time is then determined by the quadrature
\beq\label{quadrature}
\int_{a=1}^a \frac{d\tilde a}{\tilde a h(\tilde a)}=H_0(t-t_0).
\eeq

Note that in the particular case $\Omega_2=\Omega_4=0$ the modified Friedmann equation \rf{geneqH} and the cubic polynomial \rf{cube} have been explored in \cite{Sushkov:2012}. In the general case a detailed and systematic analysis of Eqs. \rf{geneqH} and \rf{cube} has been performed in Ref. \cite{StaSusVol:2016}, where was found three different branches of ghost-free solutions. In \cite{StaSusVol:2016} these solutions have been labeled as S, A, and B ones. Among them the most interesting and important with physical point of view is the S solution, which describes a universe with the standard late time dynamic dominated by the $\Lambda$-term, radiation and dust. In the case $\Omega_2=0$ ($k=0$) the S solution represents screening properties at early times of universe evolution, when the matter effects are totally screened and the universe expands with a constant Hubble rate determined by the coupling parameter $\eta$, so that $H\approx\sqrt{1/9\eta}$. Moreover, in Ref. \cite{StaSusVol:2019} it has been shown that the S solution provides the screening mechanism such that the anisotropies within the Bianchi I homogeneous spacetime model are screened at
early time by the scalar charge (see also \cite{Galeev_etal:2021}). 

%

\section{Cosmological scenarios}
For given model parameters $\zeta$ and $\Omega_i$, Eq. \rf{geneqH} completely determines the scale factor $a(t)$ and hence the whole cosmological evolution of the Universe. It is necessary noticing that the parameters are not independent. 
Actually, at $t=t_0$ one has $a_0=1$ and $h_0=1$, then Eq. \rf{geneqH} reduces to
\beq\label{constr}
1=\Omega_{0}-\Omega_2+\Omega_{3}+\Omega_{4} 
+\frac{\Omega_{6}\big(1-3\zeta(3+\Omega_2)\big)}{\big(1-3\zeta(1+\Omega_2)\big)^2}.
\eeq
The latter represents a constraint relating values of parameters $\Omega_{0}$, $\Omega_2$, $\Omega_{3}$, $\Omega_{4}$, and $\Omega_{6}$ at the present time. 
For practical purposes, it will be convenient to rewrite the constraint \rf{constr} as follows
\beq\label{omega6}
\Omega_6=\frac{\big(1-3\zeta(1+\Omega_2)\big)^2}{1-3\zeta(3+\Omega_2)}\,
(1-\Omega_0+\Omega_2-\Omega_3-\Omega_4).
\eeq 
Thus, one has five independent parameters $\zeta$, $\Omega_0$, $\Omega_2$, $\Omega_3$, $\Omega_4$ with additional requirements: $\zeta\ge 0$, $\Omega_0\ge 0$, $\Omega_3\ge 0$, $\Omega_4\ge 0$, and $\Omega_6\ge 0$.

Below we consider several cosmological models with different sets of parameters.


\subsection{The case $\zeta=0$ and $\Omega_0=\Omega_3=\Omega_4=0$}
First of all, for the sake of completeness, let us briefly discuss the simplest case with $\zeta=0$, i.e. the non-minimal coupling is absent, and $\Omega_0=\Omega_3=\Omega_4=0$, i.e. the cosmological constant, radiation, and non-relativistic matter are absent. In this case Eq. \rf{geneqH} reduces to the simple form 
\beq
h^2=-\frac{\Omega_2}{a^2}+\frac{\Omega_6}{a^6},
\eeq 
with the constraint $\Omega_6=1+\Omega_2$, which describes a cosmological evolution of an ordinary massless scalar field in the Friedmann universe. 
It is obvious that at early times, when $a\to 0$, one has $h^2\approx \Omega_6/a^{-6} \to \infty$, that is an initial cosmological singularity. The later evolution essentially depends on the sign of $\Omega_2$, i.e. on the spatial curvature of the universe. As usually, in the case of zero spatial curvature, when $k=0$ and $\Omega_2=0$, one has an open model with $h^2= \Omega_6/a^2 \to 0$ as $a\to\infty$. In the case of negative spatial curvature, when $k=-1$ and $\Omega_2<0$, one has an open model with $h^2\approx |\Omega_2|/a^2 \to 0$ as $a\to\infty$. In case the spatial curvature is positive, i.e. $k=+1$ and $\Omega_2>0$, the scale factor $a$ achieves its maximum value $a_{max}=\max(a(t))$ at $t=t_{turn}$. The moment $t=t_{turn}$ is a turning point in the universe evolution, when the expansion stage is changing to contraction one. The value of $a_{max}$ can be determined from the condition $h_{turn}^2=-\frac{\Omega_2}{a_{max}^2}+\frac{\Omega_6}{a_{max}^6}=0$, so that
\beq\label{amax_zeta=0}
a_{max}^2=\left(\frac{\Omega_6}{\Omega_2}\right)^{1/2}=
\left(1+\frac{1}{\Omega_2}\right)^{1/2}.
\eeq
Taking into account that $\Omega_2\ll 1$, we obtain the following estimation: $a_{max}^2\approx \Omega_2^{-1/2}\gg 1$, or $\textrm{a}_{max}^2\approx \textrm{a}_0^2 \Omega_2^{-1/2}\gg \textrm{a}_0^2$.  

The graphical illustration of the properties discussed above is given in Fig. \ref{fig1}.
\begin{widetext}
\onecolumngrid	
\begin{figure}[t]
	\begin{center}
		\includegraphics[scale=0.35]{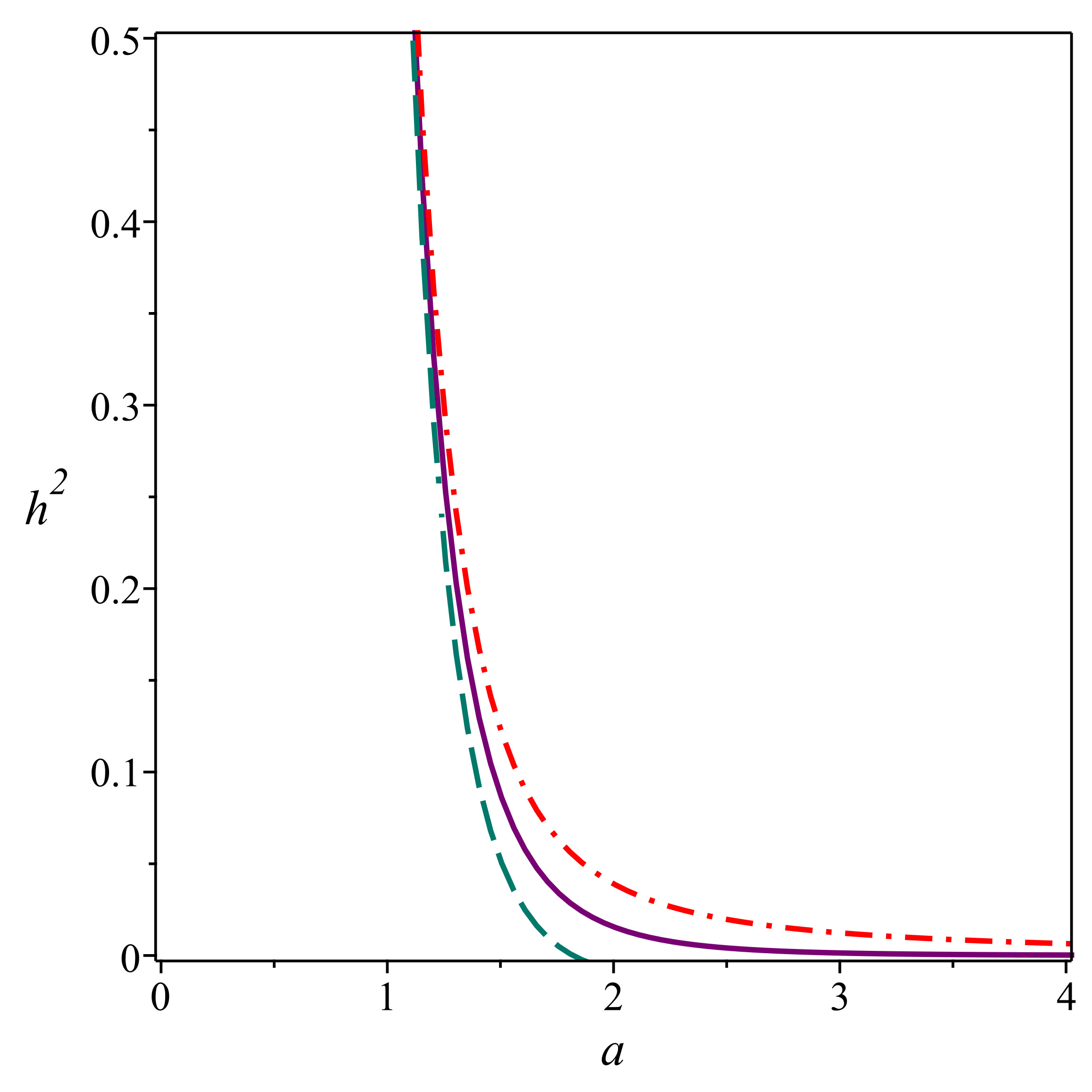}
		\hfil
		\includegraphics[scale=0.35]{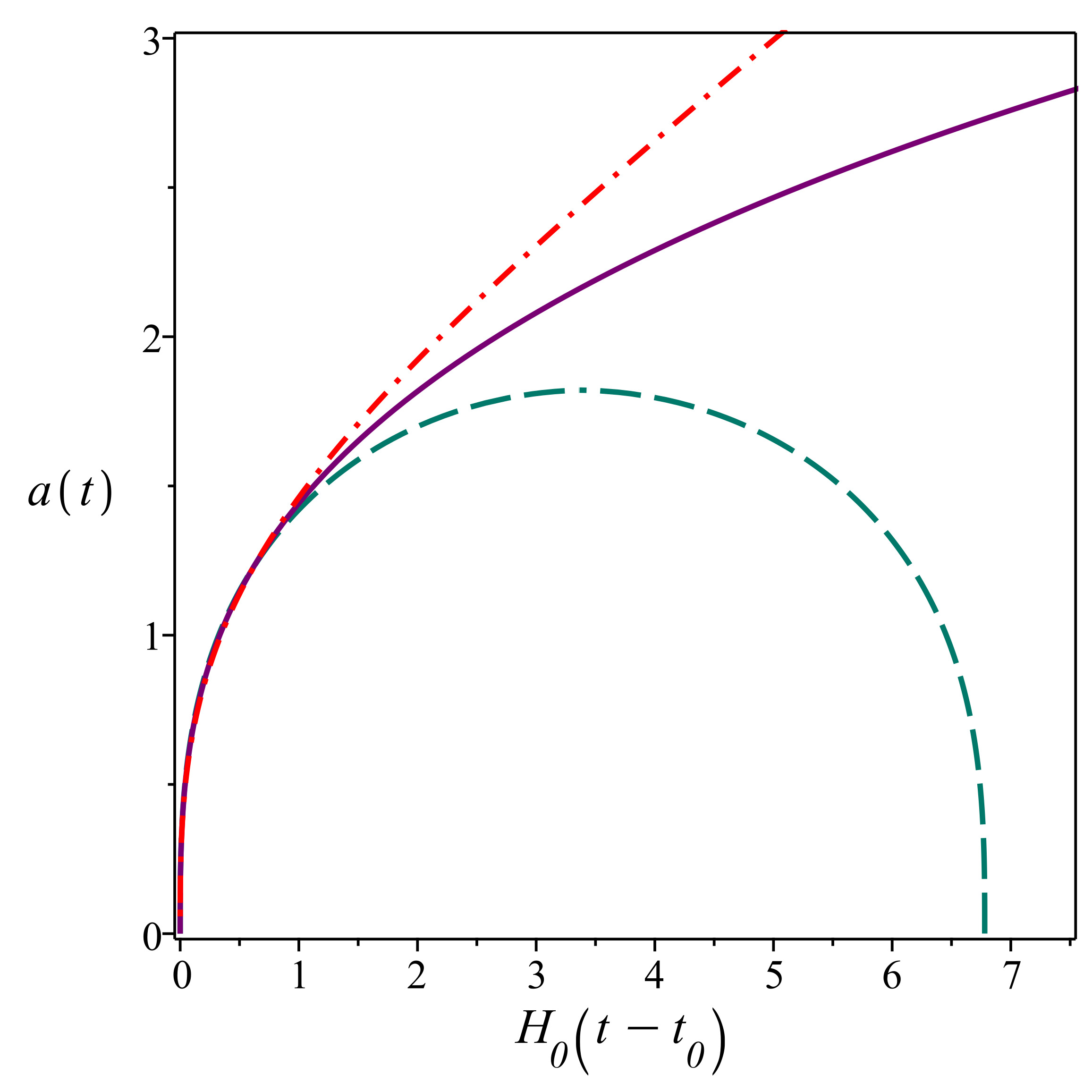}
	\end{center}	
	\caption{{The case $\zeta=0$ and $\Omega_0=\Omega_3=\Omega_4=0$.} {\em Left panel}: Plots of $h^2$ versus $a$. {\em Right panel}: Plots of $a(t)$. Here solid lines correspond to $k=\Omega_2=0$; dot-dash lines: $k=-1$ and $\Omega_2=-0.1$; and dash lines: $k=+1$ and $\Omega_2=0.1$.
		\label{fig1}}
\end{figure}
\end{widetext}

\subsection{The case $\zeta\not =0$ and $\Omega_0=\Omega_3=\Omega_4=0$}
\label{sec_Omega0=0}
Now, let us consider the model with non-minimal derivative coupling $\zeta\not=0$, while $\Omega_0=\Omega_3=\Omega_4=0$, i.e. the cosmological constant, radiation, and non-relativistic matter are still absent. Hereafter, it will be convenient to consider separately cosmological models with different spatial curvature, $k=0,-1,+1$.  
  
\subsubsection{Zero spatial curvature: $k=0$ and $\Omega_2=0$}
In this case Eq. \rf{geneqH} reads
\beq\label{k=0}
h^2=
\frac{\Omega_{6}(1-9\zeta h^2)}
{a^6(1-3\zeta h^2)^2},
\eeq
and the constraint \rf{omega6} yields
\beq
\Omega_6=\frac{(1-3\zeta)^2}{1-9\zeta},
\eeq
hence one has the only free parameter $\zeta$ in this case. The equation \rf{k=0} has been already studied in great details in the literature (see, for example, Refs. \cite{Sushkov:2009, SarSus:2010, Sushkov:2012, StaSusVol:2016, StaSusVol:2019}). As is well known, the non-minimal derivative coupling essentially changes the character of cosmological evolution at early stages. Namely, in the limit $a\to0$, the Hubble parameter $h$ has the following asymptotic behavior:   
\beq\label{as_k=0}
h^2=\frac{1}{9\zeta} +O(a^6).
\eeq
Therefore, at early cosmological times, $t\to -\infty$, one has an `eternal' inflation with the quasi-De Sitter behavior of the scale factor:
$a(t)\propto e^{H_\eta t}$, where $H_\eta=1/\sqrt{9\eta}$. It is important to notice that the primary inflationary epoch is only driven by non-minimal derivative or kinetic coupling between the scalar field and curvature without introducing any fine-tuned potential, and so one can call this epoch as a kinetic inflation.
At late times, the $\zeta$-terms in Eq. \rf{k=0} become negligibly small, and one has the asymptotic 
\beq
h^2=\frac{\Omega_6}{a^6} +O(a^{-9}), 
\eeq
the same that in case $\zeta=0$, when the universe evolution is driven only by the scalar field with the following behavior of the scale factor $a(t)\propto t^{1/3}$.

It is worth noting again that one needs no fine-tuning potential to provide the epochs change. The epoch of kinetic inflation is changed by the scalar field epoch once the $\zeta$-terms in Eq. \rf{k=0} become negligibly small. Mathematically, it means that the sign of second derivative $\ddot a$ changes its sign. Therefore, one can define a moment $t_*$ of epoch change as $\ddot a(t_*)=0$. In Fig. \ref{fig2} we represent graphs of $h^2$ versus $a$ for different values of $\zeta$ and show the typical dependence of scale factor $a(t)$ on time $t$.  

\begin{widetext}
\onecolumngrid
\begin{figure}[t]
	\begin{center}
		\includegraphics[width=0.4\linewidth]{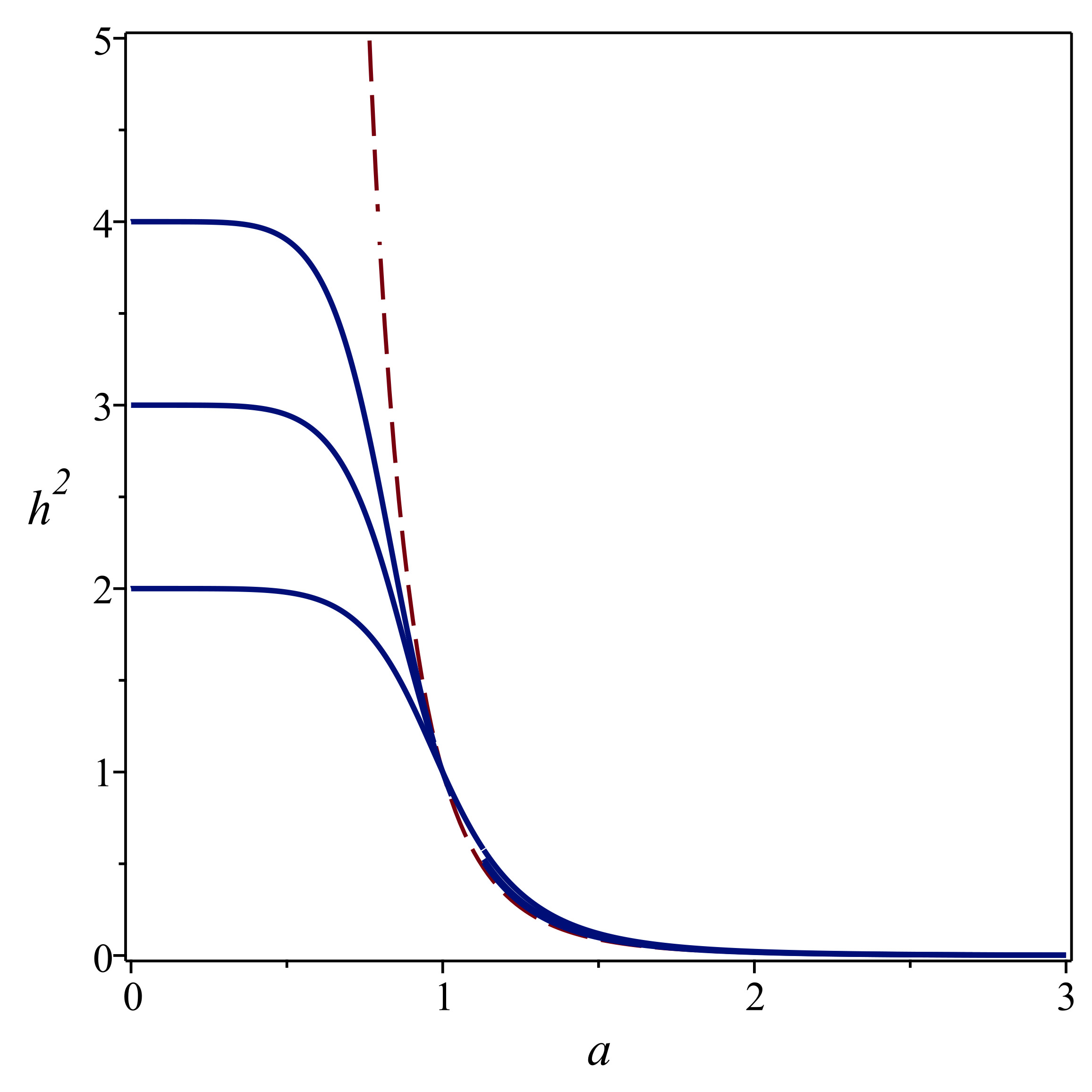}
		\hfil
		\includegraphics[width=0.4\linewidth]{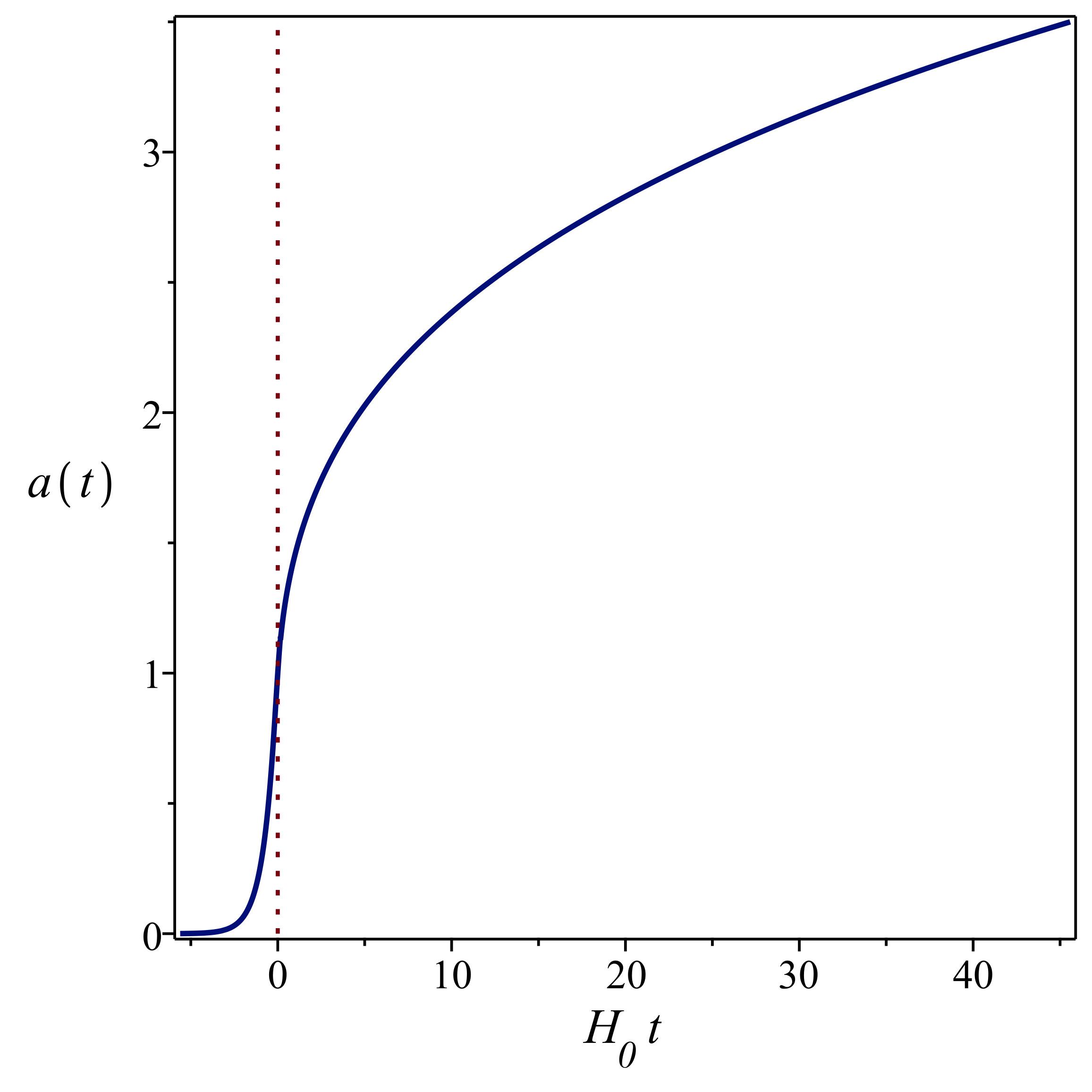}
	\end{center}	
	\caption{The case of zero spatial curvature $k=\Omega_2=0$, and also $\zeta\not=0$ (non-zero derivative coupling), and $\Omega_0=\Omega_3=\Omega_4=0$ (cosmological constant, radiation, and non-relativistic matter are absent). {\em Left panel}: Plots of $h^2$ versus $a$ (solid curves) are given for $\zeta=1/18$, $1/27$, $1/36$ from bottom to top. The dash line corresponds to $\zeta=0$. {\em Right panel}: The plot of $a(t)$ is given for $\zeta=1/18$. The vertical dot straight line separates two cosmological epochs: on the left side -- eternal kinetic inflation era, on the right side -- scalar field era.
		\label{fig2}}
\end{figure}
\end{widetext}

\subsubsection{Negative spatial curvature: $k=-1$ and $\Omega_2<0$}
In this case Eq. \rf{geneqH} reads
\beq\label{k=-1}
h^2=
\frac{\bar\Omega_2}{a^2}+
\frac{\Omega_{6}\big(1-3\zeta (3h^2-\frac{\bar\Omega_2}{a^2})\big)}
{a^6\big(1-3\zeta (h^2-\frac{\bar\Omega_2}{a^2})\big)^2},
\eeq
where $\bar\Omega_2=-\Omega_2>0$,
and the constraint \rf{omega6} yields
\beq\label{omega6_k=-1}
\Omega_6=\frac{\big(1-3\zeta(1-\bar\Omega_2)\big)^2}{1-3\zeta(3-\bar\Omega_2)}\,
(1-\bar\Omega_2).
\eeq
The relation \rf{omega6_k=-1} means that we have two free parameters $\zeta$ and $\bar\Omega_2$.

At early times, in the limit $a\to 0$, the asymptotic solution of Eq. \rf{k=-1} is as follows:
\beq\label{as_k=-1}
h^2=\frac{\bar\Omega_2}{3 a^2} + \left(\frac{1}{9\zeta}+ 
\frac{8\zeta\bar\Omega_2^3}{27\Omega_6}\right) 
+\frac{4\bar\Omega_2^2 (3\Omega_6 -8\zeta^2\bar\Omega_2^3)}{81\Omega_6^2}\, a^2
+O(a^3). 
\eeq
One can see that in distinct to the case $k=0$ ($\Omega_2=0$) with the asymptotic \rf{as_k=0}, 
the Hubble parameter $h$ has a {\em singular} behavior at $a\to 0$, so that $h^2\approx \bar\Omega_2/3 a^{-2}\to \infty$. As $a$ increases, the first term in the asymptotic \rf{as_k=-1} decreases and at $a^2 \ge a_{*}^2= \left(\frac{1}{3\zeta\bar\Omega_2}+\frac{8\zeta\bar\Omega_2}{9\Omega_6}\right)^{-1}$  it becomes negligible with respect to the second term. One can call the stage with $0<a<a_*$ as a {\em post-singularity era}. As the scale factor $a$ grows further, the behavior of Hubble parameter is determined by the second term in \rf{as_k=-1}, so that $h^2\approx h_{dS}^2= \frac{1}{9\zeta}+\frac{8\zeta\bar\Omega_2^3}{27\Omega_6}$. This stage with $a_*<a<a_{**}$ can be called as a {\em quasi-de Sitter era} with the de Sitter parameter $h_{dS}$. In Fig. \ref{fig3}, where the graphical representation for $h^2$ versus the scale factor $a$ is shown, one can see that plots $h^2(a)$ have a plateau at $h^2\approx h_{dS}^2$. This plateau is the more flat the less is values of $\bar\Omega_2$, and in the limit $\bar\Omega_2\to 0$ the plot of $h^2(a)$ coincides with that given in Fig. \ref{fig2} for the case $\Omega_2=0$.
At the end of quasi-de Sitter era the universe enters the last era of the late evolution,
which does not depend on $\zeta$ and is determined by the following late-time asymptotic at $a\to\infty$ \cite{StaSusVol:2016}:
\beq
h^2=\frac{\bar\Omega_2}{a^2} +O\left(a^{-6}\right).
\eeq

The dependence of scale factor $a$ on the cosmic time $t$ can be found from the quadrature \rf{quadrature}. In particular, the corresponding behavior of $a(t)$ near the singularity, where $h^2\approx \bar\Omega_2/3 a^{-2}$, is the following: 
\beq
a(t)\approx \sqrt{\frac{\bar\Omega_2}{3}}H_0(t-t_s),
\eeq
where $t_s$ is a moment of singularity. 
The example of $a(t)$ is shown in Fig. \ref{fig3} (right panel).

\begin{widetext}
\onecolumngrid
\begin{figure}[t]
	\begin{center}
		\includegraphics[width=0.4\linewidth]{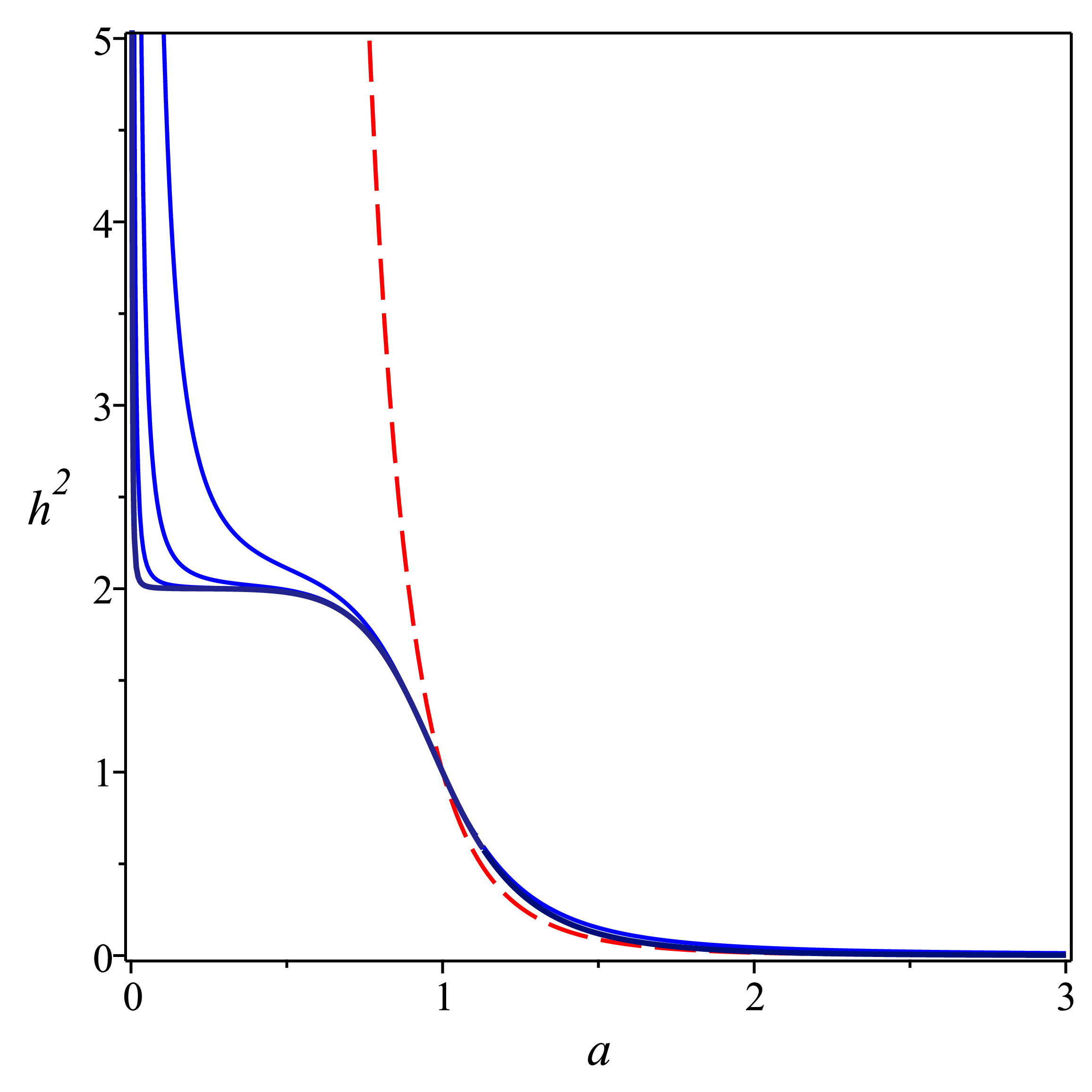}
		\hfil
		\includegraphics[width=0.4\linewidth]{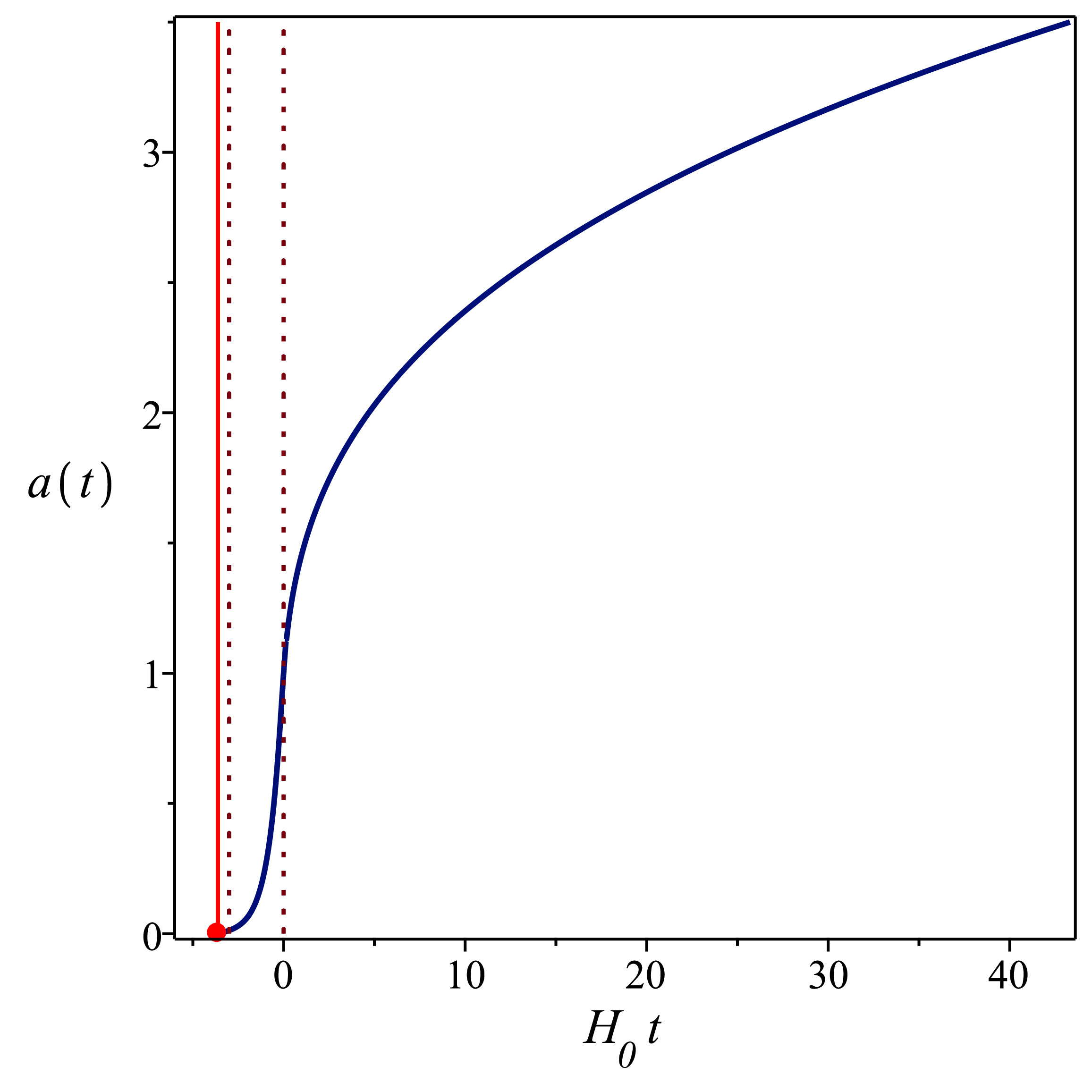}
	\end{center}	
	\caption{The case of negative spatial curvature $k=-1$ and $\Omega_2<0$, and also $\zeta\not=0$ (non-zero derivative coupling), and $\Omega_0=\Omega_3=\Omega_4=0$ (cosmological constant, radiation, and non-relativistic matter are absent). {\em Left panel}: Plots of $h^2$ versus $a$ (solid curves) are given for $\zeta=1/18$ and $\Omega_2=-0.0001$, $-0.001$, $-0.01$, $-0.1$ from bottom to top. The dash line corresponds to $\zeta=0$ and $\Omega_2=-0.0001$. {\em Right panel}: The plot of $a(t)$ is given for $\zeta=1/18$ and $\Omega_2=-0.001$. The vertical straight lines mark different epochs of the universe evolution: the solid line marks a moment of  initial singularity, the interval between two vertical lines corresponds to the kinetic inflation era, and on the right side of dot line one has the era of power-law expansion.  
		\label{fig3}}
\end{figure}
\end{widetext}

\subsubsection{Positive spatial curvature: $k=+1$ and $\Omega_2>0$}
In this case Eq. \rf{geneqH} reads
\beq\label{k=+1}
h^2=
-\frac{\Omega_2}{a^2}+
\frac{\Omega_{6}\big(1-3\zeta (3h^2+\frac{\Omega_2}{a^2})\big)}
{a^6\big(1-3\zeta (h^2+\frac{\Omega_2}{a^2})\big)^2},
\eeq
and the constraint \rf{omega6} yields
\beq\label{omega6_k=+1}
\Omega_6=\frac{\big(1-3\zeta(1+\Omega_2)\big)^2}{1-3\zeta(3+\Omega_2)}\,
(1+\Omega_2).
\eeq
At early times, in the limit $a\to 0$, the asymptotic solution of Eq. \rf{k=+1} is as follows:
\beq\label{as_k=+1}
h^2=-\frac{\Omega_2}{3 a^2} + \left(\frac{1}{9\zeta}- 
\frac{8\zeta\Omega_2^3}{27\Omega_6}\right) 
+\frac{4\Omega_2^2 (3\Omega_6 +8\zeta^2\Omega_2^3)}{81\Omega_6^2}\, a^2
+O(a^3). 
\eeq
One can see that the asymptotic behavior of $h^2$ given by \rf{as_k=+1} essentially differs from those given by \rf{as_k=0} and \rf{as_k=-1}.
Namely, since the first term in \rf{as_k=+1} is negative, at some small minimal value of $a=a_{min}$ the value of $h^2$ becomes to be zero. Neglecting the third term in \rf{as_k=+1}, one has 
\beq
a_{min}^2\approx 3\zeta\Omega_2\,\left(1-\frac{8\zeta^2\Omega_2^2}{3\Omega_6}\right)^{-1}.
\eeq
Supposing that the spatial curvature is small, so that $\zeta\Omega_2\ll 1$, we can estimate $a_{min}$ as follows: $a_{min}^2\approx 3\zeta\Omega_2\ll 1$. It must be recalled that the moment $t_B$ when the Hubble parameter $h$, or $\dot a$, equals to zero is a turning point in the universe evolution. Moreover, since at $t=t_B$ the scale factor $a$ achieves its minimal value, $a_{min}=\min(a(t))=a(t_B)$, the moment $t_B$ is a {\em bounce}, when the stage of contraction is changing to expansion one. It is interesting that we can estimate the minimal size of the universe. Actually, returning in the relation $a_{min}^2=3\zeta\Omega_2$ to the dimensional values $\zeta=\eta H_0^2=\ell^2 H_0^2$, $\Omega_2=1/(\textrm{a}_0^2 H_0^2)$, and $a=\textrm{a}/\textrm{a}_0$, we obtain 
\beq 
\textrm{a}_{min}=\sqrt{3}\,\ell,
\eeq
where $\ell$ is the characteristic scale of nonminimal derivative coupling. Thus, the minimal size of the universe is of order of $\ell$. 

\begin{figure}[t]
	\begin{center}
		\includegraphics[width=0.9\linewidth]{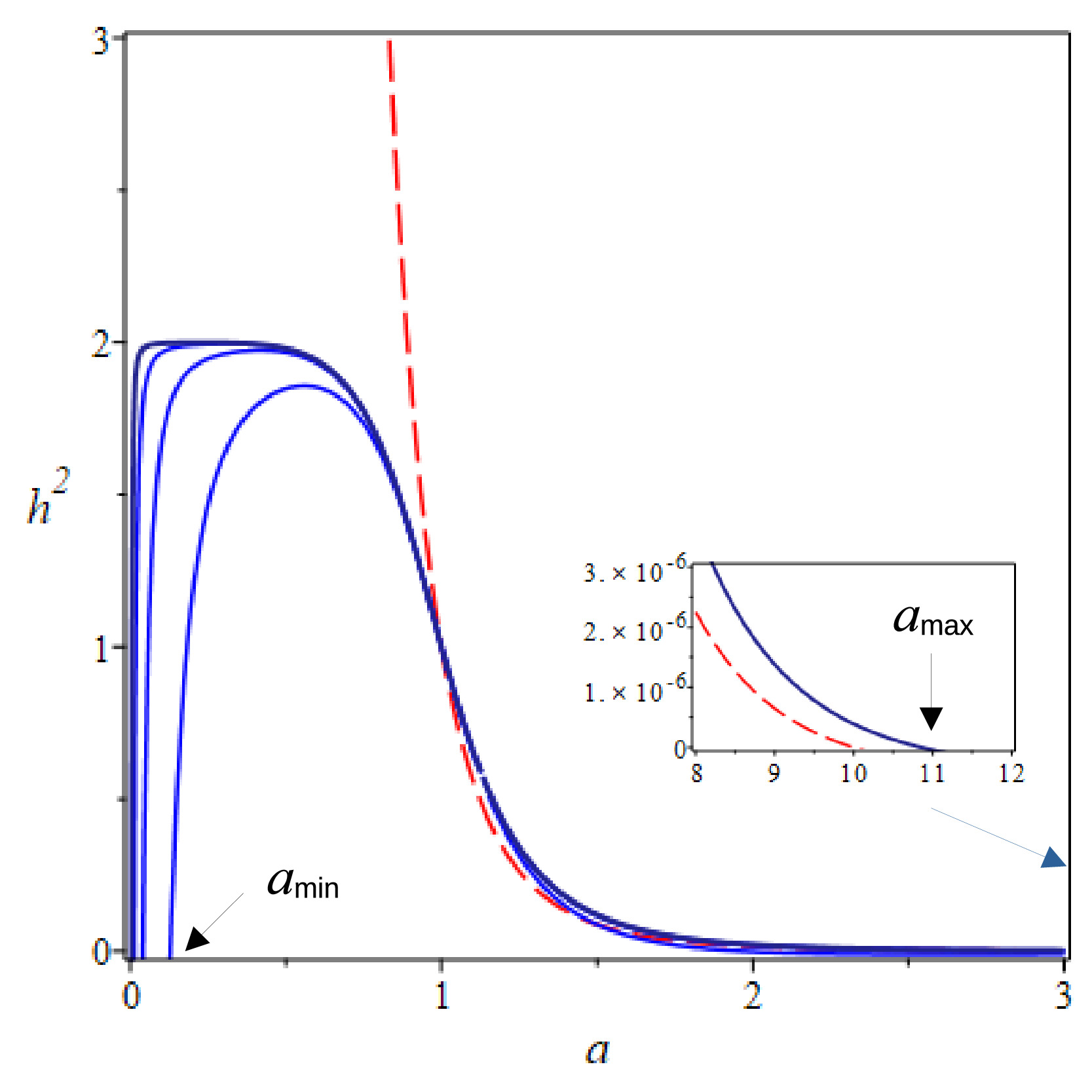}
	\end{center}	
	\caption{The case of positive spatial curvature $k=+1$ and $\Omega_2>0$, and also $\zeta\not=0$ (non-zero derivative coupling), and $\Omega_0=\Omega_3=\Omega_4=0$ (cosmological constant, radiation, and non-relativistic matter are absent). Plots of $h^2$ versus $a$ (solid curves) are given for $\zeta=1/18$ and $\Omega_2=0.0001$, $0.001$, $0.01$, $0.1$ from top to bottom. The dash line corresponds to $\zeta=0$ and $\Omega_2=0.0001$. 
	\label{fig4}}
\end{figure}

Analogously to the case of negative spatial curvature, the first term in the asymptotic \rf{as_k=+1} decreases as $a$ increases and becomes negligible comparing with the second term. As long as the second term in \rf{as_k=+1} is dominating, the Hubble parameter is approximately constant, so that $h^2\approx h_{dS}^2 = \frac{1}{9\zeta}-\frac{8\zeta\Omega_2^3}{27\Omega_6}$, and the universe goes through the {\em quasi-de Sitter phase} with the Hubble parameter $h_{dS}$.
In Fig. \ref{fig4} one can see that plots of $h^2(a)$ have a plateau at $h^2\approx h_{dS}^2$. This plateau is the more flat the less is values of $\Omega_2$, and in the limit $\Omega_2\to 0$ the plot of $h^2(a)$ coincides with that given in Fig. \ref{fig2} for the case $\Omega_2=0$.

At the end of quasi-de Sitter era the universe enters the last era of the late evolution. Characterizing this era, it is necessary to stress that $h^2$ turns out being zero at some value of $a_{max}=\max(a)$. Substituting $h^2=0$ into Eq. \rf{k=+1} yields:
\beq\label{amax_k=+1}
0=
-\Omega_2+
\frac{\Omega_{6}}
{a_{max}^4}\left(1- \frac{3\zeta\Omega_2}{a_{max}^2}\right)^{-1},
\eeq
Taking into account that $3\zeta\Omega_2 \ll 0$, one can obtain
\beq
a_{max}^2\approx \left(\frac{\Omega_6}{\Omega_2}\right)^{1/2}\,
\left(1+ \frac{3\zeta\Omega_2^{3/2}}{2\Omega_6^{1/2}}\right).
\eeq
Comparing with the value of $a_{max}$ obtained for $\zeta=0$ (see \rf{amax_zeta=0}), one can conclude that the maximal value of the scale factor $a_{max}$ is slightly greater in case $\zeta\not=0$.
  
\begin{figure}[t]
	\begin{center}
		\includegraphics[width=1.0\linewidth]{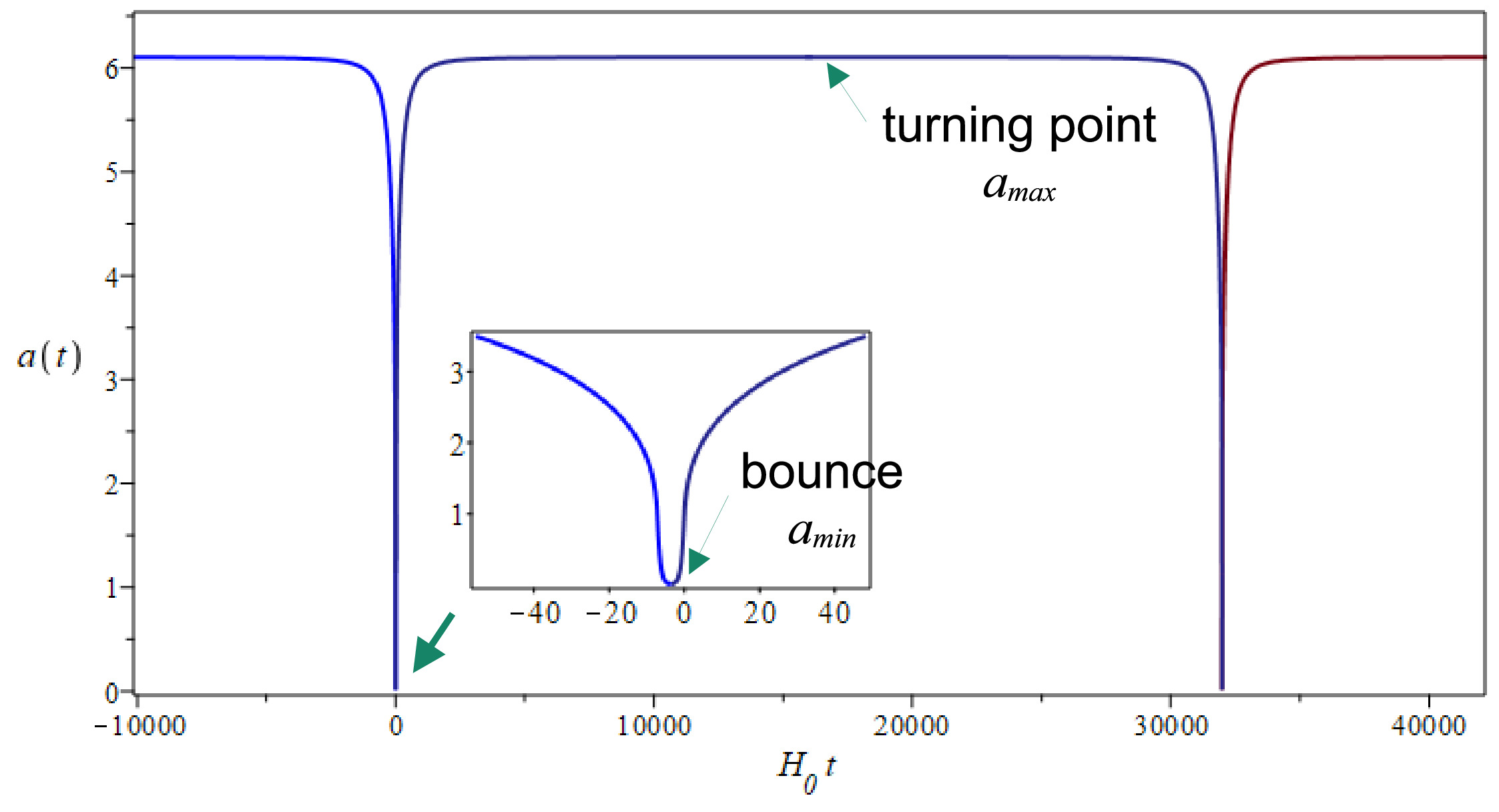}
	\end{center}	
	\caption{The case of positive spatial curvature $k=+1$ and $\Omega_2>0$, and also $\zeta\not=0$ (non-zero derivative coupling), and $\Omega_0=\Omega_3=\Omega_4=0$ (cosmological constant, radiation, and non-relativistic matter are absent). The plot of $a(t)$ is given for $\zeta=1/18$ and $\Omega_2=0.001$. On this plot one entire cycle of cyclic cosmological evolution is presented. 
		\label{fig5}}
\end{figure}

The dependence of scale factor $a$ on the cosmic time $t$ is found from the quadrature \rf{quadrature}. In particular, taking into account that near the bounce 
\beq
h^2 \approx -\frac{\Omega_2}{3 a^2} + \frac{1}{9\zeta},
\eeq
where we suppose $\zeta\Omega_2 \ll 1$, one obtains the explicit behavior of $a(t)$: 
\beq
a^2(t)\approx 3\zeta\Omega_2\, \cosh^2 \frac{H_0(t-t_B)}{\sqrt{9\zeta}},
\eeq
where $t_B$ is a moment of bounce. 
The example of $a(t)$ is shown in Fig. \ref{fig5}. One can see that the scale factor $a(t)$ has a cyclic behavior. Each cycle begins at a bounce moment when $a(t)$ achieves its minimal value $a_{min}$. Then the universe comes to a quasi-de Sitter stage with $a(t)\propto e^{h_{dS}t}$. After the end of quasi-de Sitter era the universe enters a stage of slow power-law expansion, which stops when the scale factor achieves its maximal value $a_{max}$. Further, the universe begins contracting, and its evolution goes in reverse order up to a next bounce moment. Therefore, we have a {\em cyclic scenario} of cosmological evolution.    

\subsection{The case $\Omega_0\not=0$ and $\Omega_3=\Omega_4=0$}
Now let us disscus the role of the cosmological constant supposing $\Omega_0\not=0$, while, as before, we will assume that $\Omega_3=\Omega_4=0$, i.e. radiation and non-relativistic matter are absent. 
\beq\label{Omega0_not=0}
h^2=\Omega_{0}
-\frac{\Omega_2}{a^2} 
+\frac{\Omega_{6}\big(1-3\zeta (3h^2+\frac{\Omega_2}{a^2})\big)}
{a^6 \big(1-3\zeta (h^2 +\frac{\Omega_2}{a^2})\big)^2}.
\eeq
The constraint \rf{omega6} now yields
\beq
\Omega_6=\frac{\big(1-3\zeta(1+\Omega_2)\big)^2}{1-3\zeta(3+\Omega_2)}\,
(1-\Omega_0+\Omega_2),
\eeq 
thus one has three free parameters $\zeta$, $\Omega_0$, and $\Omega_2$.

\begin{figure}[t]
	\begin{center}
		\includegraphics[width=1.0\linewidth]{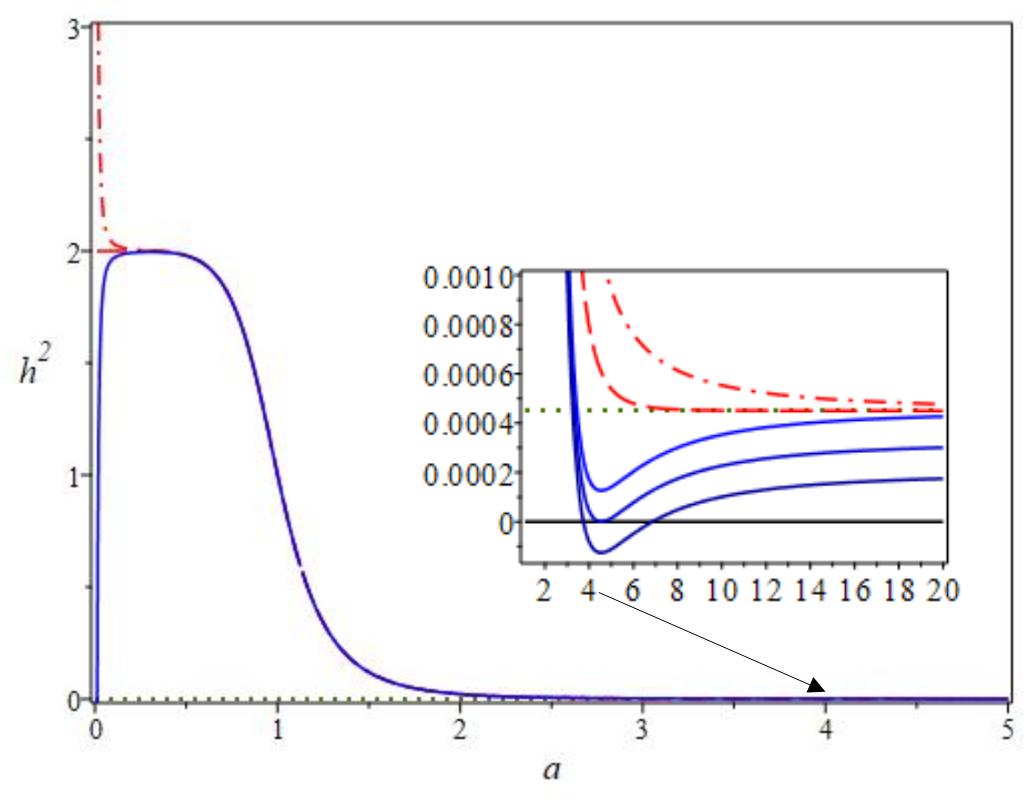}
	\end{center}	
	\caption{The case $\Omega_0\not=0$ (non-zero cosmological constant),  $\zeta\not=0$ (non-zero derivative coupling), and $\Omega_3=\Omega_4=0$ (radiation and non-relativistic matter are absent). {\em Main panel}: Plots of $h^2$ versus $a$ are given for $\zeta=1/18$, $\Omega_0=4.5\times 10^{-4}$, and $\Omega_2=0$ (red dash line), $\Omega_2=-0.01$ (red dot-dash line), $\Omega_2=0.01$ (blue solid line). {\em Auxiliary panel}: An illustration of qualitatively different behavior of $h^2$ depending on the value of $\Omega_0$. Blue solid lines are plots of $h^2$ versus $a$ given for $\Omega_0=4.5; 3.26; 2\times 10^{-4}$ from top to bottom. The dot line shows the asymptotic $h^2\approx \Omega_0=4.5\times 10^{-4}$.  
		\label{fig6}}
\end{figure}

At early times, in the limit $a\to 0$, the asymptotic solution of Eq. \rf{Omega0_not=0} reads:
\bea
h^2 &=& -\frac{\Omega_2}{3 a^2} + \left(\frac{1}{9\zeta} - 
\frac{8\zeta\Omega_2^3}{27\Omega_6}\right) 
\nonumber\\
&&+\frac{4\bar\Omega_2^2 (3\Omega_6 +8\zeta^2\Omega_2^3 +9\zeta\Omega_0\Omega_6)}{81\Omega_6^2}\, a^2
+O(a^3).\ \
\label{as_Omega0_not=0} 
\eea
It is important to stress here that first two major terms in the asymptotic \rf{as_Omega0_not=0} do not contain the cosmological constant $\Omega_0$ and coincide with those given by asymptotics \rf{k=0}, \rf{k=-1}, and \rf{k=+1} ($k=0,-1,+1$, respectively). Following Ref. \cite{StaSusVol:2016}, we may say that the cosmological constant is screened at the early stage and makes no contribution to the universe evolution which, therefore, is the same as described in Sec. \ref{sec_Omega0=0} for the case $\Omega_0=0$. Briefly, the possible scenarios of the early time universe evolution are the following:

\begin{itemize}
 \item[(i)] in the case $\Omega_2=0$ ($k=0$) at early cosmological times, $t\to-\infty$, one has an {\em eternal kinetic inflation} with the quasi-de Sitter behavior of the scale factor: $a(t)\propto e^{h_{dS} (H_0t)}$, where $h_{dS}^2=1/9\zeta$; 
 
 \item[(ii)] in the case $\Omega_2<0$ ($k=-1$) one has an {\em initial singularity} at $a\to 0$, so that $h^2\approx |\Omega_2|/3 a^{-2}\to \infty$.
Then, after a short post-singularity era the universe enters a {\em primary quasi-de Sitter epoch} with the de Sitter parameter $h_{dS}^2= \frac{1}{9\zeta}+\frac{8\zeta|\Omega_2|^3}{27\Omega_6}$. 

 \item[(iii)] in the case $\Omega_2>0$ ($k=+1$) one has a {\em bounce} at $t=t_B$, when the Hubble parameter turns to zero at some small minimal value of $a=a_{min}$, where $a_{min}^2\approx 3\zeta\Omega_2$. Shortly after the bounce the universe enters a {\em primary quasi-de Sitter epoch} with the de Sitter parameter $h_{dS}^2= \frac{1}{9\zeta}-\frac{8\zeta\Omega_2^3}{27\Omega_6}$. 
\end{itemize}
An illustration of these scenarios is given in Fig. \ref{fig6}. 

An asymptotic solution of Eq. \rf{Omega0_not=0} at large values of $a$ is as follows
\beq\label{as_Omega_0_not=0}
h^2=\Omega_0 -\frac{\Omega_2}{a^2} +\frac{\Omega_6(1-9\zeta\Omega_0)}{(1-3\zeta\Omega_0)^2}\,\frac{1}{a^6} +O(a^{-8}).
\eeq
In the case $\Omega_2\le 0$, i.e. when $k=0$ or $k=-1$, it is obvious that the value of $h^2$ given by \rf{as_Omega_0_not=0} is monotonically decreasing to $\Omega_0$, i.e. $h^2\approx \Omega_0$ at $a\to\infty$ (see Fig. \ref{fig6}). 
In the case $\Omega_2>0$ ($k=+1$) a possible scenario is more complicated. 
Since the second term in \rf{as_Omega_0_not=0} is negative when $\Omega_2>0$, the behavior of $h^2$ is now not monotonic, so that $h^2$ has a minimum 
\beq\label{h_min}
h^2_{min}=\Omega_0 -\frac{\Omega_2}{a_{*}^2} +\frac{\Omega_6(1-9\zeta\Omega_0)}{(1-3\zeta\Omega_0)^2}\,\frac{1}{a_{*}^6},
\eeq 
where $a_{*}$ can be found from the extremum condition $d(h^2)/da=0$ as
\beq
a_*^4=\frac{3\Omega_6(1-9\zeta\Omega_0)}{\Omega_2(1-3\zeta\Omega_0)^2}.
\eeq
Note that, depending on a relation between parameters $\zeta$, $\Omega_0$, and $\Omega_2$, one has $h_{min}^2>0$ or $h_{min}^2\le0$. 
In case $h^2_{min}>0$ the Hubble parameter $h$ achieves its minimal value $h_{min}$ at $a=a_*$ and then starts growing, so that $h^2\to\Omega_0$ at $a\to\infty$. 
In case $h^2_{min}\le 0$ the square of Hubble parameter $h$ becomes equal zero at some value of the scale factor $a=a_{max}$ at $t=t_{turn}$. The moment $t_{turn}$ is a turning point in the universe evolution, when the expansion stage is changing to contraction one.

\begin{figure}[t]
		\begin{center}
			\includegraphics[width=1.0\linewidth]{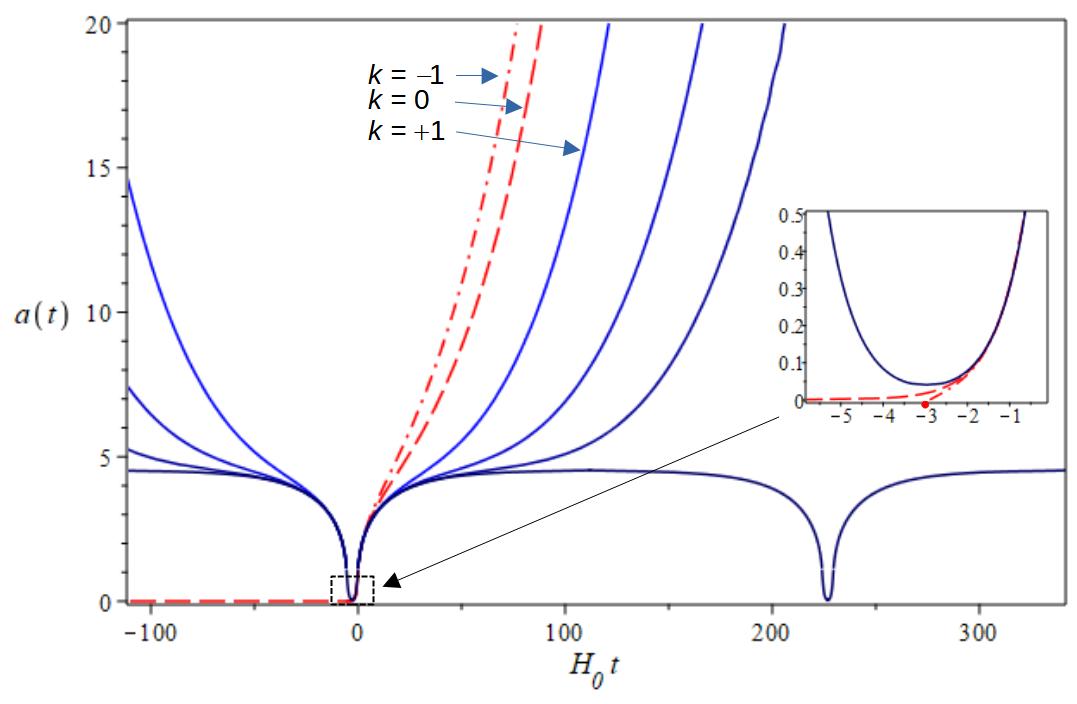}
		\end{center}	
		\caption{The case $\Omega_0\not=0$ (non-zero cosmological constant),  $\zeta\not=0$ (non-zero derivative coupling), and $\Omega_3=\Omega_4=0$ (radiation and non-relativistic matter are absent). Plots of $a(t)$ are given for $\zeta=1/18$. {\em Main panel}: The red dash line corresponds to $\Omega_2=0$ (zero spatial curvature, $k=0$), and $\Omega_0=4.5\times 10^{-4}$. The red dot-dash line corresponds to $\Omega_2=-0.01$ (negative spatial curvature, $k=-1$), and $\Omega_0=4.5\times 10^{-4}$. Blue solid lines correspond to $\Omega_2=0.01$ (positive spatial curvature, $k=+1$), and $\Omega_0=4.5; 3.5; 3.3; 3.26\times 10^{-4}$ from top to bottom. {\em Auxiliary panel}: An illustration of qualitatively different behavior at small values of $a(t)$ depending on the value of spatial curvature. One has (i) an eternal kinetic inflation if $k=0$ (red dash line); (ii) an initial singularity if $k=-1$ (red dot-dash line); (iii) a bounce if $k=+1$ (blue solid line). 
			\label{fig7}}
	\end{figure}

Summarizing, we obtain two possible scenarios of late-time evolution of the universe:
\begin{itemize}
	\item[(i)] In the case $\Omega_2\le 0$, at the late stage of evolution the universe enters a {\em secondary inflation epoch} with $h^2=\Omega_0$, i.e. $H=H_\Lambda=\sqrt{\Lambda/3}$. In the case $\Omega_2>0$ one has the same asymptotic if the value of $h_{min}^2$ given by \rf{h_min} is positive.
	
	\item[(ii)] In the case $\Omega_2>0$ and $h^2_{min}\le 0$, there is a turning point in the universe evolution, when the expansion stage is changing to contraction one. In this case one has a {\em cyclic scenario} of the universe evolution.
\end{itemize}
All possible scenarios of cosmological evolution in the case $\zeta\not=0$ and $\Omega_0\not=0$ are shown in Fig. \ref{fig7}.
%

\subsection{The general case}
The standard scenario of cosmological inflation suggests that the energy density of matter filling the universe is very slowly varying with time. The energy density of ordinary (baryon) matter does not possess that property. Instead, one supposes that the inflationary stage of the universe evolution is driven by hypothetical inflaton field, while the ordinary matter is absent on this stage and appears only at the end of inflation due to the reheating process when the inflaton is transforming into ordinary matter.  

The kinetic inflation discussing in this paper is based on the mechanism which differs from the slow-roll inflation. Therefore, one has no reasons to assume {\em a priory}\, that ordinary matter is absent during the kinetic inflationary stage. In this section we will analyze the most general cosmological model with non-minimal derivative coupling: 
\beq \label{general}
h^2=\Omega_{0}
-\frac{\Omega_2}{a^2} 
+\frac{\Omega_3}{a^3} 
+\frac{\Omega_4}{a^4}
+\frac{\Omega_{6}\big(1-3\zeta (3h^2+\frac{\Omega_2}{a^2})\big)}
{a^6 \big(1-3\zeta (h^2 +\frac{\Omega_2}{a^2})\big)^2},
\eeq
supposing that $\Omega_3\not=0$ and $\Omega_4\not=0$, that is non-relativistic and relativistic components of matter are present at all stages of the universe evolution. 

For small values of $a$ one can obtain the solution of Eq. \rf{general} as a series in powers of $1/a$:
\bea
h^2
&&=\frac{\mu_4}{a^4} +\frac{\mu_3}{a^3} +\frac{\mu_2}{a^2} +\frac{\mu_1}{a} +\mu_0 +\dots 
\nonumber\\
&&=\frac{\Omega_4}{a^4}+\frac{\Omega_3}{a^3}
-\frac{1}{a^2}\left(\Omega_2+\frac{\Omega_6}{\zeta\Omega_4}\right)
+\frac{1}{a}\,\frac{\Omega_3\Omega_6}{\zeta\Omega_4^2}
\nonumber\\
&&~~~+\left(\Omega_0 +\frac{2\Omega_2\Omega_6}{3\zeta\Omega_4^2} 
-\frac{\Omega_3^2\Omega_6}{\zeta\Omega_4^3} -\frac{\Omega_6^2}{\zeta^2\Omega_4^3}\right) 
\nonumber\\
&&~~~+O(a^2).
\label{early_as_gen} 
\eea
It is seen that in the limit $a\to 0$, the function $h^2$ has a clear singular behavior, such that $h^2\approx\Omega_4/a^4\to\infty$. Near the singularity the formula \rf{early_as_gen} represents an approximate solution if the series is convergent, i.e. at least $\mu_4/a^4>\mu^3/a^3>\mu_2/a^2>\mu_1/a>\mu_0$. In particular, taking into account that $\mu_4/a^4>\mu_0$ and $\mu_4=\Omega_4$, $\mu_0\sim1/\zeta^2\Omega_4^3$, we obtain $a<a_*\approx\zeta^{1/2}\Omega_4$, where $a_*$ is the convergence radius of the series \rf{early_as_gen}. Since values of $\zeta$ and $\Omega_4$ could be arbitrary small, the value $a_*$ is also arbitrary small. On the other hand, since terms in Eq. \rf{early_as_gen} have different signs, the behavior of $h^2$ at $a_*<a\ll 1$ could be rather complicated and messy. In particular, we can expect that $h^2$ can change sign and vanish, so that $h^2=0$, in the region $a_*<a\ll 1$. To describe a behavior of $h^2$ at $a\ll 1$ in more details, we use a graphical representation of the function $h^2$ versus $a$ for $a\ll 1$. The dependence of $h^2(a)$ for small values of $a$ is illustrated in Fig. \ref{fig8} separately for $\Omega_2=0$ (zero spatial curvature), $\Omega_2<0$ (negative spatial curvature), and $\Omega_2>0$ (positive spatial curvature). Though the general asymptotic is $h^2\approx \Omega_4/a^4\to\infty$ at $a\to 0$, it is seen that in {\em all} cases there exist non-monotonic solutions such that $h^2$ becomes zero, $h^2=0$, at some $a=a_{min}$. These points are a {\em bounce}.

The main conclusion which one can extract from the numerical analysis is the following: Analyzing the role of radiation and non-relativistic matter in the universe evolution in the theory of gravity with non-minimal derivative coupling, we found that for {\em all} types of spatial geometry of the homogeneous universe, namely, $k=-1$, $\Omega_2<0$ (negative spatial curvature), $k=0$, $\Omega_2=0$ (zero spatial curvature), $k=+1$, $\Omega_2>0$ (positive spatial curvature), there exists a wide domain of parameters $\Omega_3$ and $\Omega_4$ such that the squared Hubble parameter $h^2$ becomes zero at $a_{min}$, where $a_*<a_{min}\ll 1$. 
The moment $t_B$ when the Hubble parameter $h$, or $\dot a$, equals to zero is a turning point in the universe evolution. Moreover, since at $t=t_B$ the scale factor $a$ achieves its minimal value, $a_{min}=\min(a(t))=a(t_B)$, the moment $t_B$ is a {\em bounce}, when the stage of contraction is changing to expansion one.
%
\begin{widetext}
\onecolumngrid
\begin{figure}[h]
	\begin{minipage}[h]{0.31\linewidth}
		\center{\includegraphics[width=1\linewidth]{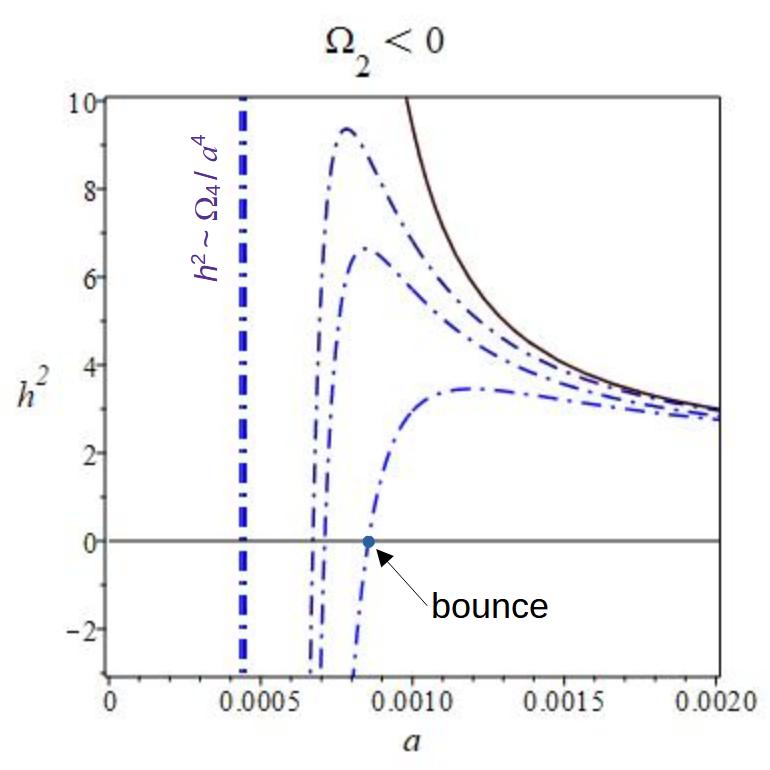}}
	\end{minipage}
	\begin{minipage}[h]{0.31\linewidth}
		\center{\includegraphics[width=1\linewidth]{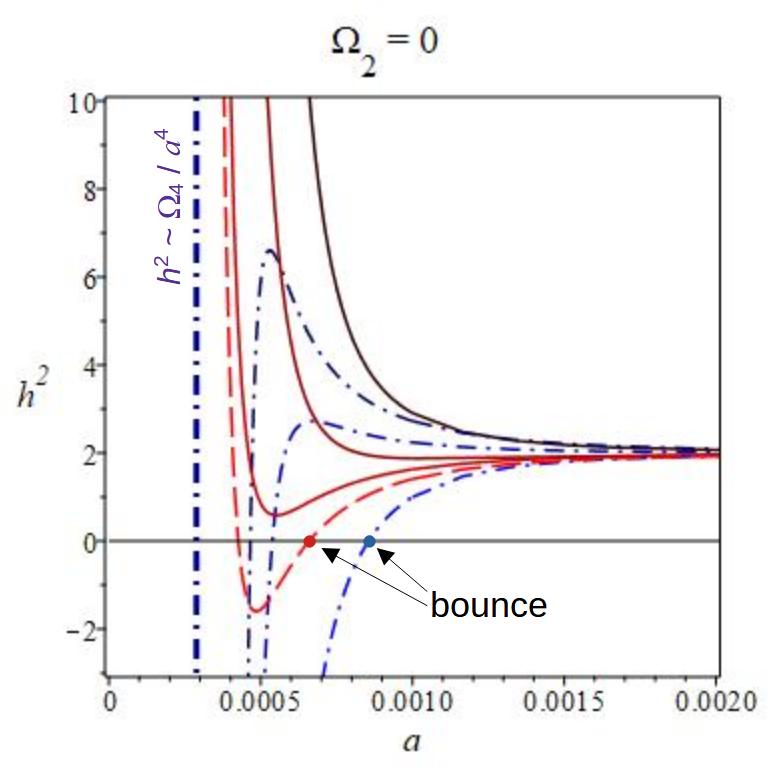}}
	\end{minipage}
	\begin{minipage}[h]{0.31\linewidth}
		\center{\includegraphics[width=1\linewidth]{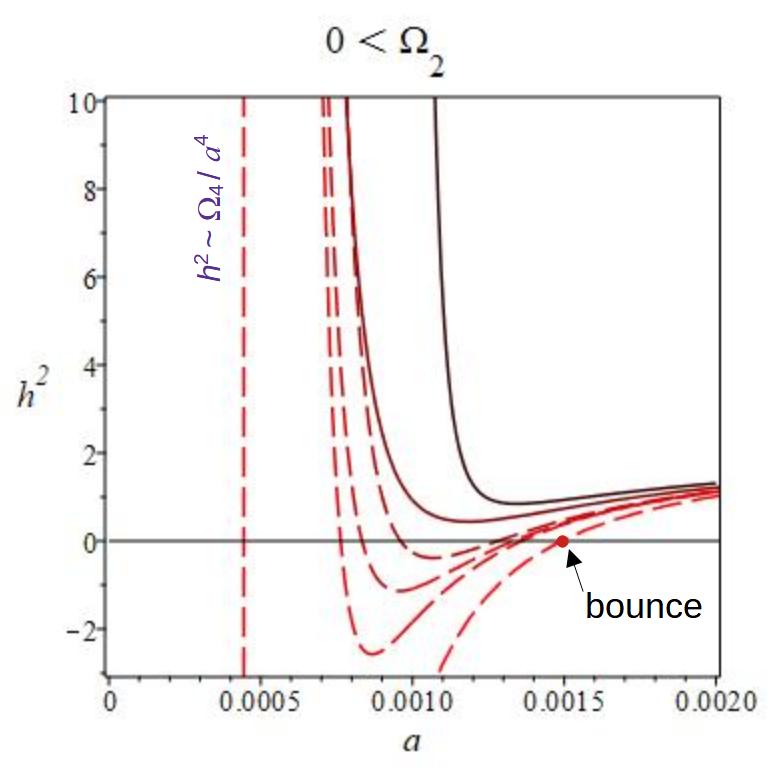}}
	\end{minipage}
\caption{The general case with $\zeta\not=0$, $\Omega_0\not=0$, $\Omega_3\not=0$, $\Omega_4\not=0$. Plots of $h^2$ versus $a$ are shown in the region of small $a$ for fixed values $\zeta=1/18$ and $\Omega_3=10^{-6}$, while $\Omega_2$ and $\Omega_4$ was varied. {\em Left panel}: $\Omega_2=-10^{-5}$ and $\Omega_4=0.0037;0.0038;0.00386;0.004$ from bottom to top. {\em Middle panel}: $\Omega_2=0$ and $\Omega_4=0.0018;0.0019;0.001905$ from bottom to top (red curves); $\Omega_4=0.0024;0.002455;0.00249;0.0026$ from bottom to top (blue curves). {\em Right panel}: $\Omega_2=10^{-5}$ and $\Omega_4=0.003807;0.003845;0.003855;0.00386;0.00387;0.0075$ from bottom to top. Note that solid curves do not cross the zero line, and hence do not give a bounce behavior, while dash curves cross zero providing the bounce condition $h^2=0$.
		\label{fig8}}
\end{figure}
\end{widetext}

An asymptotic solution of Eq. \rf{general} at large values of $a$ has the following form:
\beq\label{as_gen}
h^2=\Omega_0 -\frac{\Omega_2}{a^2} +\frac{\Omega_3}{a^3} +\frac{\Omega_4}{a^4}  +\frac{\Omega_6(1-9\zeta\Omega_0)}{(1-3\zeta\Omega_0)^2}\,\frac{1}{a^6} +O(a^{-8}).
\eeq
Comparing with the asymptotic \rf{as_Omega_0_not=0}, we can conclude that possible scenarios of late-time universe evolution coincide in the general case with those described in the previous section for the case $\Omega_0\not=0$ and $\Omega_3=\Omega_4=0$. 
Therefore,  if $\Omega_2 \le 0$, then the universe enters an epoch of accelerated expansion or a secondary inflationary epoch with $H=H_\Lambda=\sqrt{\Lambda/3}$. If $\Omega_2>0$, then the late-time universe evolution determines by the value of critical parameter $h^2_{min}$:
\beq\label{h_min2}
h^2_{min}=\Omega_0 -\frac{\Omega_2}{a_{*}^2} +\frac{\Omega_3}{a_{*}^3}
+\frac{\Omega_4}{a_{*}^4} +\frac{\Omega_6(1-9\zeta\Omega_0)}{(1-3\zeta\Omega_0)^2}\,\frac{1}{a_{*}^6},
\eeq 
where $a_{*}$ can be found from the extremum condition $d(h^2)/da=0$.
In case $h^2 _{min}>0$, at late times the universe is expanded with an acceleration so that $H=H_\Lambda=\sqrt{\Lambda/3}$, while if $h^2 _{min} \le 0$, there is a turning point in the universe evolution, when the expansion stage is changing to contraction one.

Thus,  the intermediate and late-time universe evolution is the same as in the case when $\Omega_3=\Omega_4=0$ (no radiation and non-relativistic matter). Therefore, the global dependence of $h^2(a)$ and $a(t)$ can be illustrated by Figs. \ref{fig6} and \ref{fig7}.

\section{Summary and conclusions}
In this paper we have explored in details homogeneous and isotropic cosmological solutions in the theory of gravity with non-minimal derivative coupling given by the action \rf{action}. In general, the model depends on six dimensionless parameters: the coupling parameter $\zeta$, and density parameters $\Omega_0$, $\Omega_2$, $\Omega_3$, $\Omega_4$, $\Omega_6$ (see Eqs. \rf{param1}, \rf{param2}), and a cosmological evolution is described by the modified Friedmann equation \rf{geneqH}. In the case $\zeta=0$ (no non-minimal derivative coupling) and $\Omega_6=0$ (no scalar field) one has the standard $\Lambda$CDM-model, while if $\Omega_6\not=0$ -- the $\Lambda$CDM-model with an ordinary scalar field. As is well-known, this model has an initial singularity, the same for all $k$ ($k=0,\pm1$), while its global behavior depends on $k$. The universe expands eternally if $k=0$ (zero spatial curvature) or $k=-1$ (negative spatial curvature), while in case $k=+1$ (positive spatial curvature) the universe expansion is changed to contraction, which is ended by a final singularity. 

The situation is {\em crucially} changed when the scalar field possesses non-minimal derivative coupling to the curvature, i.e. when $\zeta\not=0$. For the cosmological model with $\Omega_3=\Omega_4=0$ (no matter), we have obtained the following results: The possible scenarios of the early time universe evolution are the following:
\begin{itemize}
	\item[(i)] in the case $\Omega_2=0$ ($k=0$) at early cosmological times, $t\to-\infty$, one has an {\em eternal kinetic inflation} with the quasi-de Sitter behavior of the scale factor: $a(t)\propto e^{h_{dS} (H_0t)}$, where $h_{dS}^2=1/9\zeta$; 
	
	\item[(ii)] in the case $\Omega_2<0$ ($k=-1$) one has an {\em initial singularity} at $a\to 0$, so that $h^2\approx |\Omega_2|/3 a^{-2}\to \infty$.
	Then, after a short post-singularity era the universe enters a {\em primary quasi-de Sitter epoch} with the de Sitter parameter $h_{dS}^2= \frac{1}{9\zeta}+\frac{8\zeta|\Omega_2|^3}{27\Omega_6}$. 
	
	\item[(iii)] in the case $\Omega_2>0$ ($k=+1$) one has a {\em bounce} at $t=t_B$, when the Hubble parameter turns to zero at some small minimal value of $a=a_{min}$, where $a_{min}^2\approx 3\zeta\Omega_2$. Shortly after the bounce the universe enters a {\em primary quasi-de Sitter epoch} with the de Sitter parameter $h_{dS}^2= \frac{1}{9\zeta}-\frac{8\zeta\Omega_2^3}{27\Omega_6}$. 
\end{itemize}
The possible scenarios of the late-time universe evolution in the case $\Omega_3=\Omega_4=0$ are the following:
\begin{itemize}
	\item[(i)] In the case $\Omega_2\le 0$, at the late stage of evolution the universe enters a {\em secondary inflation epoch} with $h^2=\Omega_0$, i.e. $H=H_\Lambda=\sqrt{\Lambda/3}$. In the case $\Omega_2>0$ one has the same asymptotic if the value of $h_{min}^2$ given by \rf{h_min} is positive.
	
	\item[(ii)] In the case $\Omega_2>0$ and $h^2_{min}\le 0$, there is a turning point in the universe evolution, when the expansion stage is changing to contraction one.
\end{itemize}

In the standard scenario of slow-roll inflation one usually supposes that  ordinary matter is absent on this stage and appears only at the end of inflation due to the reheating process when the inflaton is transforming into ordinary matter. However, since the kinetic inflation discussing in this paper is based on the mechanism which differs from the slow-roll inflation, one has no reasons to assume {\em a priory}\, that ordinary matter is absent during the kinetic inflationary stage. In our work we have analyzed the most general cosmological model with non-minimal derivative coupling containing non-relativistic and relativistic components of matter at all stages of the universe evolution. As the result, we have found that 
there exists a wide domain of parameters $\Omega_3$ and $\Omega_4$ such that the squared Hubble parameter $h^2$ becomes zero at some moment $t_B$ when the scale factor $a$ achieves its minimal value, $a_{min}=\min(a(t))=a(t_B)$. This moment $t_B$ is nothing but a {\em bounce}, when the stage of contraction is changing to expansion one. It is important that the bounce is possible for {\em all} types of spatial geometry of the homogeneous universe. 
%

Concluding this paper, it is worth to enumerate once more several basic results obtained:
 \begin{itemize}
 \item  
 The cosmological constant $\Lambda$ (or $\Omega_0$) turns out to be {\em screened} at early times and makes no contribution to the universe evolution (see also Ref. \cite{StaSusVol:2016}). 
 
 \item
Depending on model parameters, there are three qualitatively different initial state of the universe: an {\em eternal kinetic inflation}, an {\em initial singularity}, and a {\em bounce}. The bounce is possible for {\em all} types of spatial geometry of the homogeneous universe. 

 \item
 For {\em all} types of spatial geometry, we found that the universe goes inevitably through the {\em primary quasi-de Sitter} (inflationary) epoch with the de Sitter parameter $h_{dS}^2= \frac{1}{9\zeta}-\frac{8\zeta\Omega_2^3}{27\Omega_6}$. For $k=0$ this epoch lasts eternally to the past, when $t\to-\infty$. When $k=-1$ or $+1$, the primary inflationary epoch starts soon after a birth of the universe from an initial singularity, or after a bounce, respectively. Here it is necessary to stress once more that the mechanism of primary or {\em kinetic} inflation is provided by non-minimal derivative coupling and needs no fine-tuned potential.  
 
\item
The kinetic inflation is driving by terms in the field equations responsible for the non-minimal derivative coupling. At early times these terms are dominating, and the cosmological evolution has the quasi-de Sitter character $a(t)\sim e^{H_\eta t}$ with $H_\eta=1/\sqrt{9\eta}$. Later on, in the course of cosmological evolution the domination of $\eta$-terms is canceled, and this leads to a {\em change} of cosmological epochs.    
 
\item
 The late-time universe evolution depends both on $k$ and $\Lambda$. In the case $k=0$ (zero spatial curvature), or $k=-1$ (negative spatial curvature), at late times the universe enters an epoch of {\em accelerated expansion} or a secondary inflationary epoch with $H=H_\Lambda=\sqrt{\Lambda/3}$. In case $k=+1$ (positive spatial curvature), the late-time universe evolution determines by the value of critical parameter $h^2_{min}$ (see Eqs. \rf{h_min} and \rf{h_min2}). In case $h^2 _{min}>0$ at late times the universe is expanded accelerated with $H=H_\Lambda=\sqrt{\Lambda/3}$, while in the case $h^2 _{min}\le 0$ there is a turning point in the universe evolution, when the expansion stage is changing to contraction one.   
 
\item  
Depending on model parameters, there are {\em cyclic scenarios} of the universe evolution with the non-singular bounce at a minimal value of the scale factor, and a turning point at the maximal one.
\end{itemize}

\section*{Acknowledgments}
This work is supported by the RSF grant No. 21-12-00130 and partially carried out in accordance with the Strategic Academic Leadership Program "Priority 2030" of the Kazan Federal University.

\end{document}